\documentclass[twocolumn]{aastex63}
\usepackage{amssymb}
\usepackage{amsmath}
\usepackage{booktabs}
\usepackage{hyperref}
\usepackage{multirow}
\usepackage{graphics}

\PassOptionsToPackage{hyphens}{url}\usepackage{hyperref}
\usepackage[hyphenbreaks]{breakurl}

\newcommand{\ltsima}{$\; \buildrel < \over \sim \;$}
\newcommand{\simlt}{\lower.5ex\hbox{\ltsima}} 
\newcommand{\gtsima}{$\; \buildrel > \over \sim \;$}
\newcommand{\simgt}{\lower.5ex\hbox{\gtsima}} 

\newcommand{\feka}{\mbox{Fe  K$\alpha$}}

\newcommand{\xmm}{{\emph{XMM-Newton}}}

\newcommand{\lum}{erg~s$^{-1}$}
\newcommand{\flux}{{erg~cm$^{-2}$~s$^{-1}$}}
\newcommand{\nh}{cm$^{-2}$}
\newcommand{\nhsym}{N_{\mbox{\scriptsize H}}}

\newcommand{\chandra}{{\emph{Chandra}}}

\newcommand{\errUD}[2]{\ensuremath{^{+#1}_{-#2}}}

\newcommand{\fexxvi}{Fe\,\textsc{xxvi}}
\newcommand{\suzaku}{{\emph{Suzaku}}}

\newcommand{\swift}{{\emph{Swift}}}
\newcommand{\logxi}{erg cm s$^{-1}$}
\newcommand{\nustar}{{\emph{NuSTAR}}}

\newcommand{\sorg}{MCG-03-58-007}

\usepackage{lineno}
\begin{document}

\title{Dramatic changes in the observed velocity of the accretion disk wind in MCG-03-58-007 are revealed by XMM-Newton and NuSTAR }
\shorttitle{Tracking the  ultrafast disk wind in MCG-03-58-007}

\shortauthors{Braito et al.}
\author{V. Braito}
\affiliation{INAF, Osservatorio Astronomico di Brera
 Via Bianchi 46
 I-23807 Merate (LC), Italy}
\affiliation{Department of Physics, Institute for Astrophysics and Computational
Sciences, The Catholic University of America, Washington, DC 20064, USA}
\author{J. N. Reeves} 
\affiliation{Department of Physics, Institute for Astrophysics and Computational
Sciences, The Catholic University of America, Washington, DC 20064, USA}
\affiliation{INAF, Osservatorio Astronomico di Brera
 Via Bianchi 46
 I-23807 Merate (LC), Italy}
 \author{G. Matzeu}
\affiliation{Department of Physics and Astronomy - DIFA, University of Bologna, Via Gobetti 93/2 - 40129 Bologna, Italy}
\author{P. Severgnini}
\affiliation{INAF - Osservatorio Astronomico di Brera, Via Brera 28, I-20121, Milano, Italy}

 \author{L. Ballo}
\affiliation{European Space Astronomy Centre (ESA/ESAC), E-28691 Villanueva de la Canada, Madrid, Spain}
\author{C. Cicone}
\affiliation{Institute of Theoretical Astrophysics, University of Oslo, P.O. Box 1029 Blindern, 0315 Oslo, Norway}
\author{R. Della Ceca}
\affiliation{INAF - Osservatorio Astronomico di Brera, Via Brera 28, I-20121, Milano, Italy}
\author{M. Giustini}
\affiliation{Centro de Astrobiolog\'ia (CSIC-INTA), Camino Bajo del Castillo s/n, Villanueva de la Ca\~nada, E-28692 Madrid, Spain}
\author{M. Sirressi}
\affiliation{Department of Astronomy, AlbaNova University Center, Stockholm University, SE-10691 Stockholm, Sweden}

\begin{abstract}
Past   X-ray observations   of the nearby Seyfert 2 \sorg\    revealed the presence of a powerful and  highly variable disk wind, where two   possible phases   outflowing with  $v_{\rm out1}/c\sim -0.07$ and $v_{\rm out2}/c\sim -0.2$ were   observed.  Multi-epoch  X-ray observations, 
covering the period from 2010 to 2018, showed that the lower velocity component   is persistent, as it was   detected in all the observations, while the faster phase outflowing with $v_{\rm out2}/c\sim -0.2$  appeared to be more sporadic. 
Here we present the analysis of a new monitoring campaign of \sorg\ performed  in May-June 2019  and consisting of   four simultaneous \xmm\ \& \nustar\ observations. We confirm that the disk wind in \sorg\ is  persistent, as it is detected in all the observations, and powerful, having a kinetic power that ranges between 0.5-10\% of the Eddington luminosity.  The  highly ionized wind (log($ \xi/{\rm erg\,cm \,s^{-1}})\sim 5$) is variable in both the opacity  and remarkably in its velocity. This is the first time where we  have observed a substantial variability of the outflowing velocity in a disk wind,   which dropped from   $v_{\rm {out}}/c\sim -0.2$ (as measured in the first three observations) to $v_{\rm {out}}/c\sim -0.074$ in just 16 days.  We conclude  that such a dramatic  and fast variability of the outflowing velocity could be    due to the acceleration of the wind, as recently  proposed by \citet{Mizumoto2021}.  Here, the faster wind,  seen in the    first three observations, is already  accelerated to $v_{\rm {out}}/c \sim -0.2$, while in the last observation our  line of sight intercepts only the   slower, pre-accelerated streamline.

\end{abstract}

\keywords{ galaxies: active -- galaxies: individual (\sorg) --  X-rays: galaxies }

\section{Introduction}\label{sec:intro}
 
It is now widely recognized that    high-velocity outflows are an almost ubiquitous and  important  phenomenon   of the central regions of Active  Galactic Nuclei (AGN; (\citealt{Tombesi2010,Tombesi2012,Gofford2013,Gofford2015}). They were initially discovered  through  observations 
 of blue-shifted absorption features from iron K-shell transitions  in the X-ray spectra of bright    AGN (PDS\,456, \citealt{Reeves2003}, PG1211+143,   \citealt{Pounds2003} and APM\,08279+5255, \citealt{Chartas2002}). The inferred  high velocities, reaching up to $\sim 0.3\,c$,  and high ionization state immediately suggested that they  originate  
from within a few hundreds of  gravitational radii  from the central black  hole (\citealt{King_Pounds2015}).  These winds are also massive and the inferred outflow rates can approach up  to a few $M_\odot$ yr$^{-1}$,  matching the typical accretion rates of AGN. This  indicates that they are  probably    linked  to  the accretion process itself  (\citealt{King2003,King2010,King_Pounds2015}) and  driven either by the radiation pressure    (\citealt{Proga2000,Proga2004,Sim2008,Sim2010b}), by  magneto-rotational forces (MHD models: \citealt{Kato2004,Kazanas2012,Fukumura2010,Fukumura2017}) or a combination of both. \\

Although they can be affected by high uncertainties, the corresponding kinetic  powers are typically of the order of a few percent of the bolometric luminosity of the AGN and thus    matching or exceeding  the conventional threshold of  $L_{\rm {KIN}}/L_\mathrm{bol} \sim0.5-5$\% for   efficient AGN  feedback on the host galaxy (\citealt{HopkinsElvis2010,DiMatteo2005}).     These ultra fast outflows may thus play an important role in shaping the   observed  AGN-host galaxy relationships like the $M-\sigma$ relation (\citealt{Magorrian1998,Ferrarese2000,Gebhardt2000}). Potentially,  they  can   simultaneously  self regulate the growth of the super massive black hole (\citealt{King_Pounds2015,Zubovas_King12,Zubovas_King16}) and drive the   massive, kpc-scales  molecular outflows      (\citealt{Cicone2014,Cicone2015,Fiore2017,Fluetsch2019,Lutz2020}),   which can sweep  away   the interstellar medium   and   quench the  star formation.

Nonetheless, the real connection between the small-scale   disk winds,  detected in the X-ray spectra,  and the large-scale molecular  outflows  is not proven yet, as there are still too few examples of powerful disk wind with deep Atacama Large Millimeter/submillimeter Array (ALMA) observations and vice versa. In this respect, the  first detections of  powerful X-ray  disk winds  in two Ultra Luminous Infrared Galaxies with   massive large-scale molecular outflows (Mrk 231; \citealt{Feruglio2015} and IRASF\,11119+3257; \citealt{Tombesi2015})  appeared to  follow the prediction for the so called blast-wave  scenario (\citealt{King2010,Faucher2012}).  According to this model, when  the nuclear disk wind   propagates and interacts with the ISM,  it produces  a  shocked wind, which  sweeps up   the  ambient gas. In particular if the Compton cooling of the shock is not  efficient, in the so called   `energy-driven'  case, the  large-scale outflow will receive a momentum boost.   
 However,  recent results   on  other  ultra fast  disk winds observed with  ALMA  show  that not all  outflows lie on the energy conserving  relation. In  the majority of the cases the  momentum rate of the large scale outflows are   well below the prediction from the energy conserving scenario and more in line with    a momentum-conserving wind (see \citealt{Veilleux2020} and references therein). We note that    for   the prototype of the fast disk winds PDS\,456 (\citealt{Bischetti2019}),   the powerful wind of I\,Zw\,1 (\citealt{ReevesBraito2019}) as well as for the powerful disk wind at the core of this work (\citealt{Sirressi2019}),  an energy conserving wind can be clearly ruled out.  This does not imply that the disk winds do not impact the star formation in the host galaxy,  but rather that they do not always couple efficiently with the ISM (\citealt{Mizumoto2019,Veilleux2020}).\\ 
 
Generally, when  we have an estimate of the  energetics of both the small and  large-scale phases of the outflow,    they are affected by large uncertainties, which   further complicates the interpretation of the results.  In the X-ray band we can improve our accuracy on the estimates of the energetics by applying self consistent disk wind models. Recently, various groups have been developing new  models for  radiatively driven  (\citealt{Sim2008,Sim2010a,Sim2010b,Hagino2015,Hagino2016,Nomura2017,Nomura2020,Mizumoto2021}) as well as MHD  winds. (\citealt{Fukumura2010,Fukumura2015,Kazanas2012}). These  models   can be now applied to the X-ray spectra of AGN winds (e. g. \citealt{ReevesBraito2019,Luminari2018}) to derive  more robust estimates of    the mass outflow rates,  the terminal velocities  and hence of the winds    energetics. \\
 
 Another important aspect that need to be considered  is that disk winds are generally extremely variable  on timescales that can be as short as a few weeks or even days (e.g. PDS~456 \citealt{Matzeu2016,Reeves2018b}), whereby they can vary in ionisation, column density ($\nhsym$) and also velocity  (e.g. PDS~456, \citealt{Reeves2018b,Matzeu2017}; IRASF\,11119+3257, \citealt{Tombesi2017}; PG\,1211+143, \citealt{pg1211};   APM\,08279+5255, \citealt{Saez2011}; IRAS~13224-3809, \citealt{Parker2018}  and \sorg, \citealt{Braito2018, Matzeu2019},   hereafter B18 and M19).   Fast variability  of the properties of the disk winds is not unexpected as disk wind simulations predict that the stream  is not a homogeneous and constant flow  (\citealt{Proga2004,Giustini2012,Dannen2020,Waters2021}).  Depending on the observation,  our line of sight could thus intercept  different clumps or streams of the winds or we could be witnessing a reaction of the wind to the changes in the  luminosity of the X-ray source.  A   direct correlation between the outflowing velocity and the luminosity of the ionizing continuum 
 has been reported for   PDS\,456 (\citealt{Matzeu2017}),  IRAS\,13224-3809  (\citealt{Chartas2018})  and APM\,08279+5255 (\citealt{Saez2011}), while a correlation between the ionisation of the disk wind and the X-ray luminosity was  also reported for IRAS\,13224-3809  by \citet{Pinto2018}.  These results   suggest  that the incident  radiation plays an important role in driving the disk winds and  when the luminosity and radiation pressure increase, a faster and/or more ionized wind is driven. \\

Here, we present  the results of a   monitoring campaign of   \sorg,  consisting of four simultaneous \xmm\
 \& \nustar\ observations taken in May-June 2019.  \sorg\ is a bright and nearby  Seyfert 2 galaxy  ($F_\mathrm{2-10\, keV}\sim 2-4 \times10^{-12}$\,erg\,cm$^{-2}$\,s$^{-1}$, $z=0.03233$; \citealt{Sirressi2019}), which  hosts one of the most powerful  and extremely variable ultra fast disk winds. 
 \sorg\ is currently optically classified as a Seyfert 2,  but the host galaxy  appears to be viewed at 37 deg (\citealt{Sirressi2019}). Therefore unless there is a clear misalignment between the central  accretion disk and the host galaxy  our line of sight could intercept the nuclear accretion disk wind  only  grazing  the outer edge of the putative torus, which could also coincide with the outer and colder part of the wind.
 \sorg\ is also another clear case where the large scale outflow, detected in the central  $\sim 4$\,kpc,    has not received  a strong momentum boost. The  CO(1-0)  emission line  seen in the ALMA observation traces a low velocity ($v_{\,\mathrm{CO}}\sim 170$ km s $^{-1}$) molecular outflow, whose kinetic power  is at least two orders of magnitude below the expected value for an energy conserving wind ($\dot E_{\rm Mol}/\dot E_{\rm X}\sim 4\times 10^{-3}$; \citealt{Sirressi2019}). 
  
 \subsection{Previous X-ray  observations of \sorg} 
  The disk wind of \sorg\  was   first discovered thanks to a deep      \suzaku\  (\citealt{Mitsuda07}) observation performed in 2010, where    two deep ($EW\sim 300$ eV)  blue-shifted absorption troughs at $E=7.4\pm 0.1$\,keV and $E=8.5\pm 0.2$\,keV  (B18) were detected. These features were  associated with  two zones of a highly ionized (log ($\xi/$\logxi)$\sim 5.5$) and high column density  ($\nhsym \sim 5-8 \times 10^{23} $\,\nh)  wind outflowing with $v_{\rm out1}/c\sim -0.1$ and $v_{\rm out2}/c\sim -0.2$. 
The presence of a persistent fast disk wind was subsequently confirmed by follow up observations performed in 2015   simultaneously with  \xmm\ \& \nustar, in 2016 with \chandra\ and in  2018 with \swift.   The first   follow-up confirmed the presence of the  slow component of the wind, but not the $\sim 8.5$ keV feature and revealed a remarkable variability of the disk wind, where  we       witnessed  an X-ray eclipse   that     lasted $\Delta t \sim 120 $ ksec. This occultation was ascribed to  a higher opacity streamline  of the wind, outflowing at $v_{\rm {out}}\sim -0.124\,c$,  which crossed our line of sight (B18). Indeed,  the  observed spectral change could not be accounted for by  variations of the  $\nhsym$  of the neutral absorber (M19).   \chandra\  (in 2016) and \swift\  (in 2018)  caught \sorg\ in a relatively  brighter state ( $F_{{\mathrm{2}-10\, keV}} \sim 4 \times 10^{-12}$ \,erg\,cm$^{-2}$\,s$^{-1}$).  These observations  further corroborate that not only the wind is persistent    but also extremely variable, with the  column density varying  in the  $\nhsym \sim 3-8  \times 10^{23}$\,\nh\ range (see \citealt{Braito2021}; hereafter B21). Interestingly, while the  \xmm\ \& \nustar\  and \chandra\  observations did not detect the   second faster phase (outflowing with $v_{\rm out2}/c\sim -0.2$),   this component was  present during the 2018 \swift\ observation.  \sorg\ is  thus  an  unique disk wind, whose variability can inform us on the structure of   the disk winds  and ultimately their driving mechanism.
  \\

 The paper is structured as follows: in \S\ref{sec:Obs_description}   we  describe the new  \xmm\ \& \nustar\ monitoring campaign  and the  data reduction, while   in  \S\ref{sec_lightcurves}  we present the variability analysis and  the overall spectral evolution. In \S\ref{spectra_std}  we report on  the initial spectral analysis, where the disk wind is modeled adopting  a  grid of  photoionized absorbers generated with the \textsc{xstar} photoionization code (\citealt{xstar}), while in \S\ref{spectra_diskw}  the disk wind is fitted with  synthetic spectra generated from the 
 the accretion disk wind model developed by \citet{Sim2008,Sim2010b}.  The disk wind energetics and variability properties are presented in \S\ref{discussion},   where we will also discuss  the overall scenario for this unique wind as inferred  from the extraordinary variability it displayed   in this monitoring. \\
 
 Throughout the paper  we assume a concordance cosmology with  $H_0=70$  km s$^{-1}$ Mpc$^{-3}$, $\Omega_{\Lambda_{0}} =0.73$    and $\Omega_m$=0.27. For the abundances we used  those of \citet{Wilms2000}.\\

\setcounter{table}{0}
 
\begin{deluxetable*}{cclclcc}[t]
\tablecaption{\label{tab:obslog} Summary of  the \xmm\ \& \nustar\ observations of \sorg\ performed in 2019: Observation number, Observatory, Start Time, Stop Time,  Instrument,  Elapsed  and Net exposure times. 
}
\tablewidth{0pt}
\tablehead{
\colhead{Obs.} & \colhead{Satellite} & \colhead{Start Date (UT Time)} & \colhead{Stop Date (UT Time)} & \colhead{Instrument}  & \colhead{Elapsed Time (ks)} & \colhead{Exposure$_{\rm(net)}$ (ks)$^{a}$}} 
\startdata
OBS 1& \xmm\ & 	2019-05-14 15:05    & 2019-05-15 11:57 &EPIC-pn  & 75.1  &  46.2 \\
 & \nustar\ & 	      2019-05-14  13:36     & 2019-05-15 11:36  &FPMA/B & 78.7  & 37.7 \\
  \\
 \hline\\
 OBS 2& \xmm\ & 	2019-05-24 03:55    & 2019-05-24 21:12 &EPIC-pn  & 62.3  &  53.8 \\
 & \nustar\ & 	        2019-05-24  03:51   & 2019-05-25 02:21 &FPMA/B  & 78.7 &  40.2 \\
\\
\hline\\
 OBS 3& \xmm\ & 	2019-05-28 15:37    & 2019-05-29 10:59 &EPIC-pn  &69.7  &  45.2 \\
 & \nustar\ & 	        2019-05-28 14:16   & 2019-05-29 11:11 &FPMA/B  & 78.9 &  37.4 \\
\\
\hline  \\
OBS 4& \xmm\ & 	2019-06-13 15:00   &  2019-06-14 09:54    &EPIC-pn  & 68.0  &  46.6  \\
 & \nustar& 	       2019-06-13 08:41  &    2019-06-14 10:41  &FPMA/B  &93.1   &40.5 (31.3$^{b}$)    \\
  \enddata
\tablenotetext{a}{The  net exposure times  are obtained after the screening of the cleaned event files for high background and dead time.}
\tablenotetext{b}{\nustar\ cleaned exposure time of the part of the observation that overlaps with the \xmm\ one.}
 \end{deluxetable*}

   \begin{figure*}
\begin{center}
\rotatebox{0}{\includegraphics[height=3.6cm]{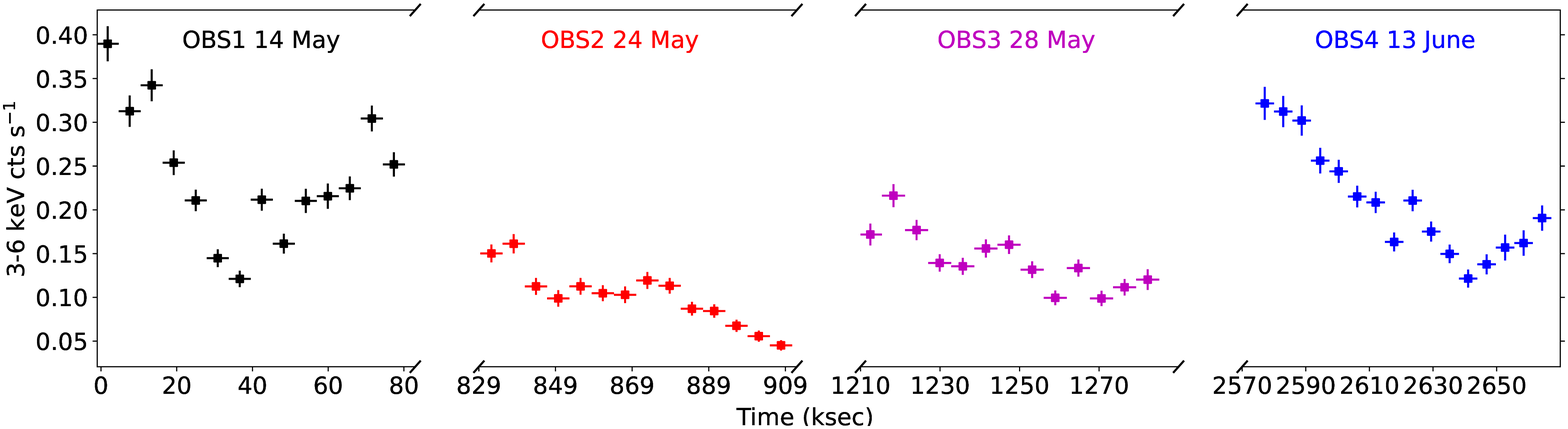}}
\rotatebox{0}{\includegraphics[height=3.6cm]{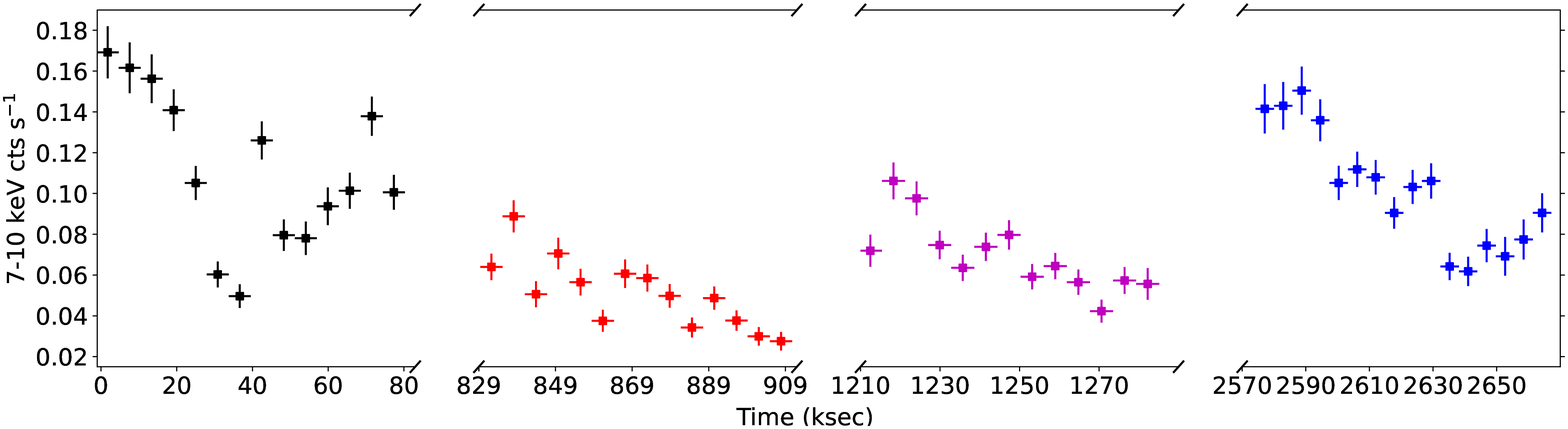}}
\rotatebox{0}{\includegraphics[height=3.6cm]{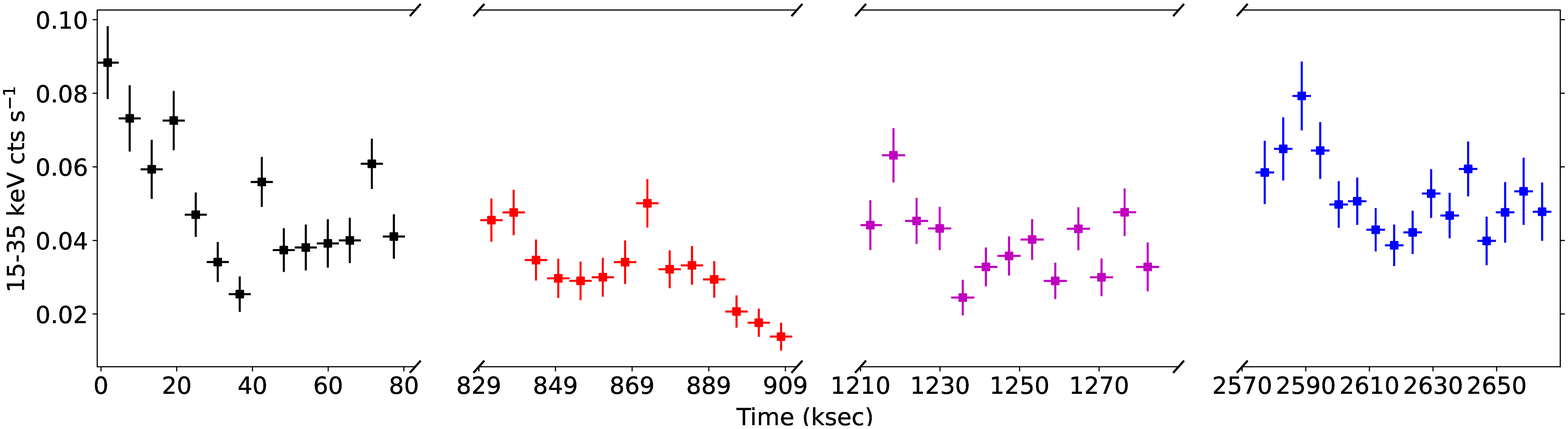}}
\rotatebox{0}{\includegraphics[height=3.6cm]{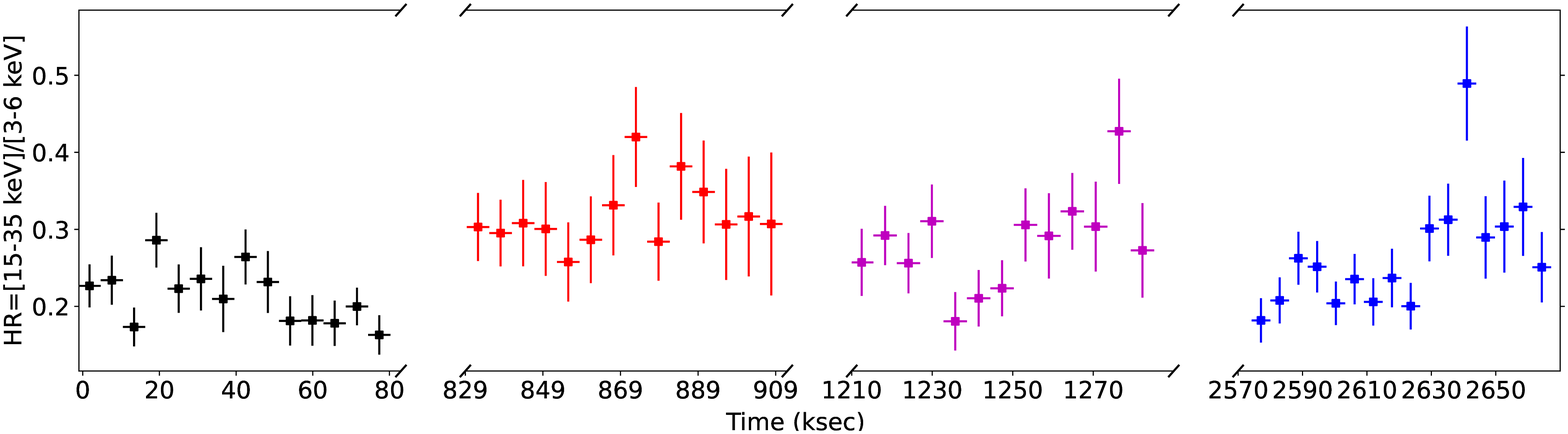}}
\end{center}
\caption{\nustar\ light curves (extracted with a bin-size of $5814$\,s) and hardness ratio  for the four observations  where: OBS1, OBS2, OBS3 and OBS4 are shown in black, red, magenta  and blue respectively.  
 From top to bottom we  show the  curves extracted  in the 3-6 keV   (upper panel), in the 7-10 keV (second panel) and in the 15-35 keV energy band (third panel). The lower panel reports the $HR_{\nustar}$ light curve, where the $HR_{\nustar}$ was defined as $CR_{\mathrm{15-35\, keV}}/CR_{\mathrm{3-6\, keV}}$. All the light curves show a degree of variability  during each of the observations, in particular in the lower energy bands. On the contrary, the $HR_{\nustar}$  is less variable on short time-scales,  but it  clearly varies between the different observations.}
\label{fig:nustar_lc}
\end{figure*}

\begin{figure*}[h]
\begin{center}
\rotatebox{0}{\includegraphics[height=3.6cm]{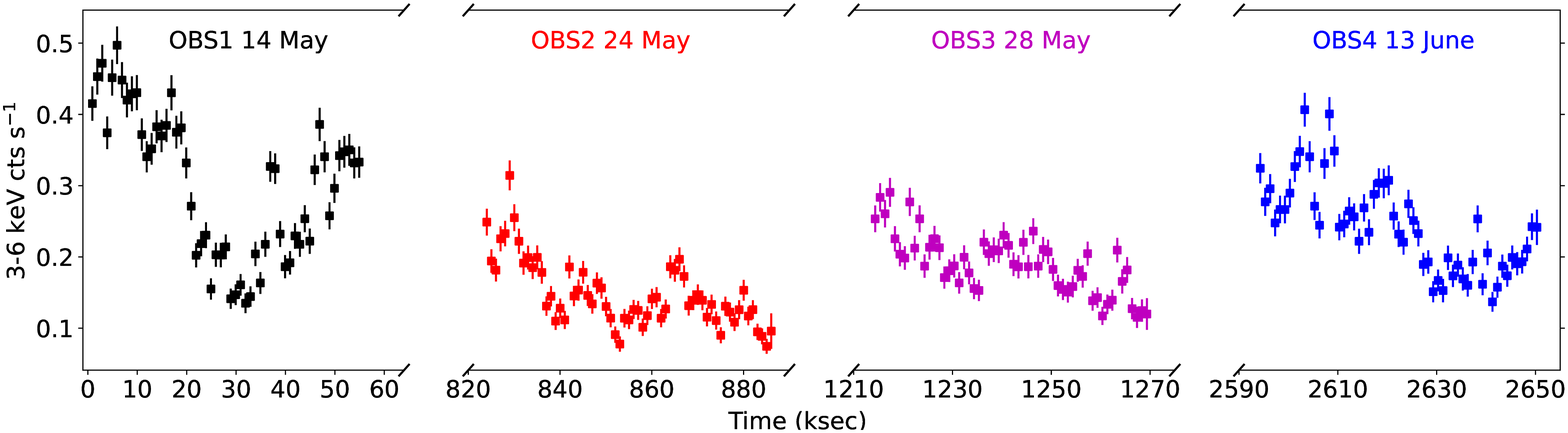}}
\rotatebox{0}{\includegraphics[height=3.6cm]{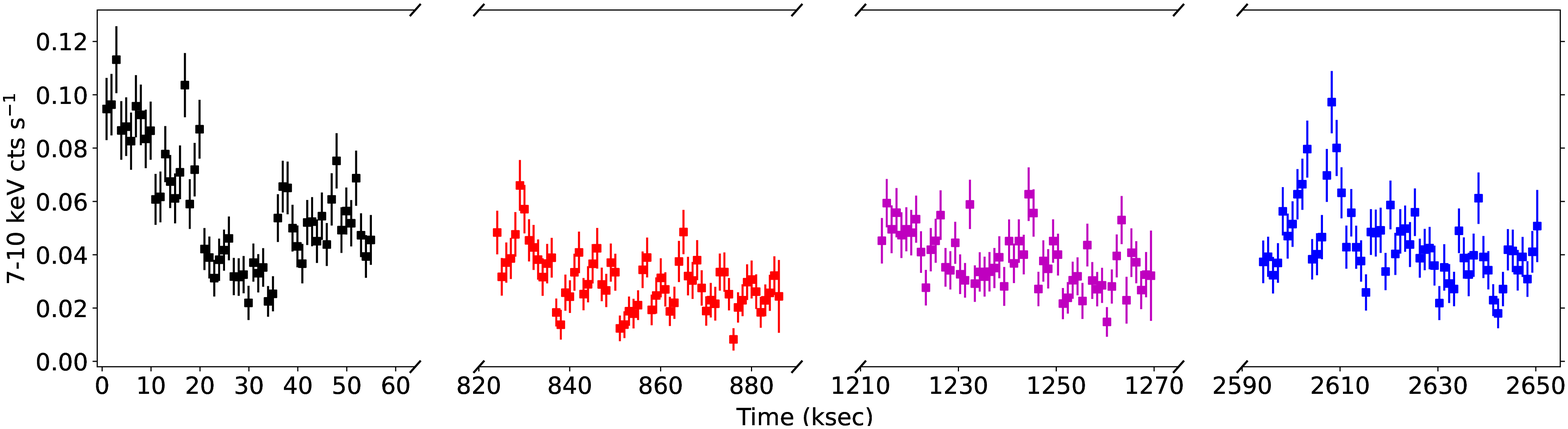}}
\rotatebox{0}{\includegraphics[height=3.6cm]{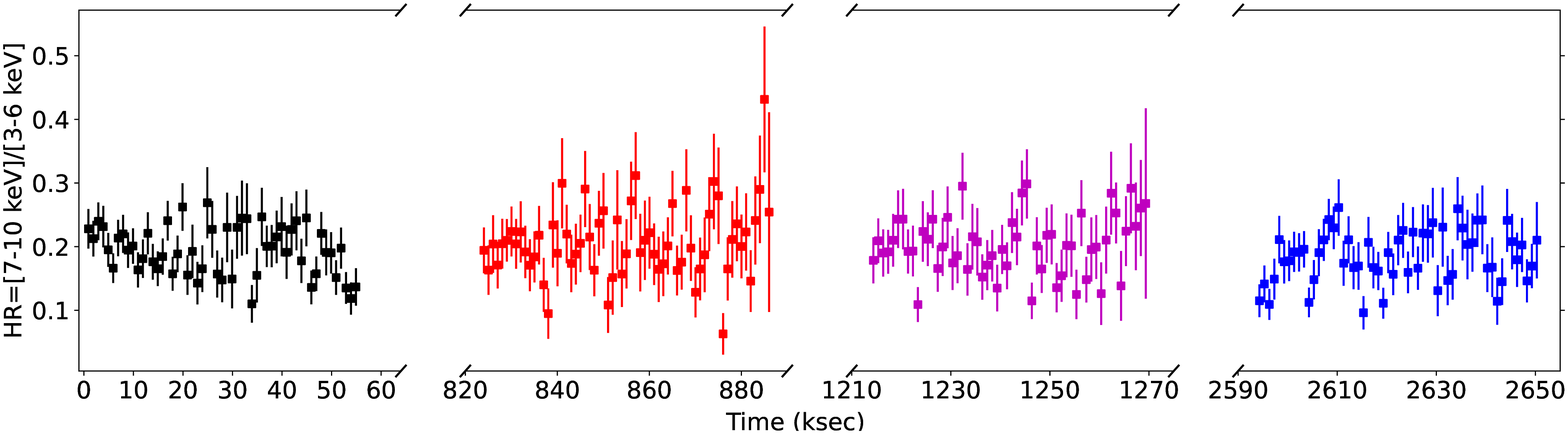}}
\end{center}
\caption{EPIC-pn light curves  and hardness ratio  for the four observations  where: OBS1, OBS2, OBS3 and OBS4 are shown in black, red, magenta  and blue   respectively. The light curves were extracted with a bin size of 1000 sec. The upper panel report the 3-6 keV light curve, while the middle panel shows the 7-10 keV light curve. \sorg\ displays a remarkable variability in each of the observations. In the lower panel we report the $HR_{XMM}$ light curve, here the   $HR_{XMM}$ was defined as $CR_{\mathrm{7-10\, keV}}/CR_{\mathrm{3-6\, keV}}$. We note that despite the strong fluxvariability, there is no clear $HR_{XMM}$ variability.  This implies that the overall spectral curvature between 3-10 keV remains roughly the same during the whole campaign.}
\label{fig:pn_lc}
\end{figure*}

\section{Observations and data reduction}\label{sec:Obs_description}
 In 2019   \sorg\  was observed four times simultaneously with \xmm\  and with the Nuclear Spectroscopic Telescope Array  (\nustar; \citealt{NUSTAR}). The observations were part of a  monitoring  campaign, designed to investigate the   variability of its fast disk wind. The observations were spaced apart by 5, 10 and 15 days, and we will refer to them as OBS1, OBS2, OBS3 and OBS4. In  Table~\,\ref{tab:obslog} we report the summary of  the four observations.  
  
  \subsection{\xmm}
 The \xmm-EPIC instruments operated in full  frame mode and with the thin filter applied. We processed and cleaned the \xmm\ data     using the Science Analysis Software  (SAS ver. 16.0.0, \citealt{SAS})  and the resulting spectra were analysed  using  standard software packages  (FTOOLS ver. 6.27.2\citealt{FTOOLS}, XSPEC ver. 12.11; \citealt{xspecref}). 
 The EPIC data were first filtered for high background, which  affected most of   the observations. The EPIC-pn  source and background  spectra  were extracted  using a
circular region  with a   radius of $32''$  and    two circular regions  with a radius of $28''$, respectively.  
We generated the response matrices and  the ancillary response files at the source position   using the SAS tasks \textit{arfgen} and \textit{rmfgen} and the latest calibration available.  For the scientific analysis    reported in this paper,  we  concentrated on the pn  data,  which have the highest signal to noise in the 2-10 keV band. The pn source spectrum was   binned to have at least 50 counts  in each energy bin.    For each of the observations,  we also extracted light curves in  the 3-6 keV and 7-10 keV energy bands. 

 \subsection{\nustar}

 All the \nustar\ observations  of \sorg\  were coordinated with  \xmm,  starting  before each \xmm\ observation and ending just after. For all the observations we considered the whole \nustar\ exposure with the exception of OBS4, where not only the \nustar\ observation started several hours before the \xmm\ one but also at the beginning of this observation \sorg\ was in a relatively brighter state  (see Fig.~\ref{fig:nustar_lc}); in this case only the portion of the \nustar\ observation that overlaps with \xmm\ was considered. We reduced the \nustar\ data following the standard procedure  using the \textsc{heasoft} task \textsc{nupipeline}   (version 0.4.6) of the \nustar\ Data Analysis Software  (\textsc{nustardas}, ver. 1.8.0). We  used the calibration files released with  the CALDB  version 20170727 and applied the standard screening criteria, where we filtered for the passages through the SAA  setting the mode    to  ``optimised"  in \textsc{nucalsaa}. For each of the Focal Plane Module (FPMA and FPMB)  the source spectra were extracted  from a circular region with a radius of $46''$,  while the background spectra were extracted  from two circular regions with a $46''$ radius located on the same detector. Light-curves in  different energy bands  were extracted from the same regions  using the \textsc{nuproducts} task. The FPMA and FPMB  background subtracted light curves  were then combined into a single one. 
 After checking   for consistency, we  combined the spectra and responses from the individual FPMA and FPMB detectors  into a single spectrum. The spectra were then binned to at least 50 counts per bin and fitted over  the 3--40 keV  energy range.

\begin{deluxetable*}{lcccccc}[b]
 \tablecaption{\label{fvar} Light curves analysis }
\tablewidth{0pt}
\tablehead{
\colhead{Obs.} & \colhead{ ${CR_{\mathrm{3-6 keV}}}$ } & \colhead{ F$_{\mathrm{VAR_{3-6}}}$}   & \colhead{ $CR _{\mathrm{7-10 keV}}$ }  & \colhead{F$_{\mathrm{VAR_{7-10}}}$} &  \colhead{$CR_{\mathrm{15-35 keV}}$ }& \colhead{ F$_{\mathrm{VAR_{15-35}}}$}   \\ 
}
\startdata
 OBS1-pn &  $0.302\pm0.003$ &$0.329\pm0.009$ & $0.058\pm0.001$ &  $0.356\pm0.021$ &-&-\\  
OBS2-pn &  $0.148\pm0.002$ &$0.300\pm0.012$ & $0.030\pm0.001$ &  $0.273\pm0.029$ &-&-\\ 
OBS3-pn &  $0.191\pm0.002$ &$0.202\pm0.012$ & $0.038\pm0.001$ &  $0.189\pm0.028$ &-&-\\ 
OBS4-pn &  $0.247\pm0.002$ &$0.252\pm0.010$ & $0.045\pm0.001$ &  $0.277\pm0.025$ &-&-\\  
 \\
 OBS1-FPM &  $0.240\pm0.004$ &$0.312\pm0.016$ & $0.111\pm0.002$ &  $0.330\pm0.023$ & $0.051\pm0.002$& $0.326\pm0.037$\\  
 OBS2-FPM &  $0.101\pm0.002$ &$0.308\pm0.023$ & $0.051\pm0.002$ &  $0.312\pm0.032$ & $0.032\pm0.001$& $0.287\pm0.044$\\  
 OBS3-FPM &  $0.142\pm0.003$ &$0.223\pm0.021$ & $0.069\pm0.002$ &  $0.231\pm0.028$ & $0.039\pm0.002$& $0.210\pm0.041$\\  
OBS4-FPM &  $0.208\pm0.003$ &$0.297\pm0.017$ & $0.102\pm0.002$ &  $0.269\pm0.023$ & $0.052\pm0.002$& $0.149\pm0.035$\\  
 \enddata
 \end{deluxetable*}

\section{The \xmm\ and \nustar\ light curves}\label{sec_lightcurves}
 In the  previous observations  \sorg\ displayed a remarkable variability  both on short and long timescale (B18, M19 and B21);  in particular in the \nustar\ observation performed in 2015,  we witnessed  a rapid occultation event due to a possible increase in the opacity of the wind (B18).  We  therefore first inspected the light curves extracted for each observation. In Fig.~\ref{fig:nustar_lc} we report the \nustar\ light curves extracted with a bin-size of 5814 sec, which corresponds to the \nustar\ orbital period. In the top three panels we show the light curves extracted in the 3-6 keV, 7-10 keV and 15-35 keV, while in the bottom panel we show the hardness  ratio (defined as  $HR_{\nustar}=CR_{\rm 15-35\, keV}/ CR_{\rm 3-6\, keV}$) as a function of time.  The three bands were chosen to sample  the spectral components that were  previously seen in the X-ray spectra of \sorg\  (see M19 and B21). The 3-6 keV samples the curvature imprinted by the fully covering  absorber, the 7-10 keV band covers the energy range  where the absorbing features ascribed to the fast wind have been detected, and  the 15-35 keV covers  the primary continuum. Similarly, we extracted the EPIC-pn light curves in the 3-6 keV and in the 7-10 keV bands with a bin-size of 1000\,s; these are shown in Fig.~\ref{fig:pn_lc}, together with the relative hardness ratio (defined as $HR_{XMM}=CR_{\rm 7-10\, keV}/ CR_{\rm 3-6\, keV}$). \\ 
 
 The inspection of the light curves shows that \sorg\ varies rapidly in all the energy bands. For example, in OBS1  all the  count rates  vary  for more than a factor of 3 on time scale as short as 10s of ksec.  However these variations do not appear to be accompanied by a fast variability of the hardness ratios.  According to   the light curves,  OBS1  and OBS2 caught \sorg\ in   the  brightest  and the   faintest state, respectively.  The $HR_{XMM}$, which tracks the 3-10 keV spectral curvature,  does not vary  between the observations (see Fig.~\ref{fig:pn_lc}, lower panel), suggesting that the overall spectral shape between 3-10 keV is similar in all the observations. Conversely    $HR_{\nustar}$, which tracks the variability  of the spectral curvature with respect to the primary continuum, varies between all the observations (see  Fig.~\ref{fig:nustar_lc}, lower panel). On short time-scale   $HR_{\nustar}$ exhibits only small fluctuations, where the only  possible change is seen  in OBS3 and  OBS4, when \sorg\ becomes marginally harder at the end of the observations. \\

 In order to quantify the source variability we calculated, for each band and detector, the `excess variance' (\citealt{Edelson2002,Vaughan2003}), which is defined as  
$\sigma^2_{\rm XS}=S^2-\overline{ \sigma_{\rm err} ^2}$, where $S^2=\frac{1}{N-1} \sum\limits_{i=1}^{N} (x_i-\overline{x})^2   $ is the sample variance, $\overline {\sigma_{\rm err} ^2}$ is the mean square error defined as  $\overline{\sigma_{\rm err} ^2} = \frac{1}{N} \sum\limits_{i=1}^{N} \sigma^2_{\rm {err},i}$, $x_i$ and $\sigma_{\rm err,i}$ are the observed values  and    the  uncertainties on each individual measurement.  From these we can then calculate the fractional root mean square (rms) variability amplitude,   defined as  $F_{\rm var} = \sqrt{\frac {\sigma^2_{\rm XS}}{\overline{x^2}}}$,
 which provides  the rms variability amplitude as percentage. In Table~\ref{fvar} we list the  arithmetic mean $\overline{x}$ and $F_{var}$  and their uncertainties   for each of the lightcurves. We note that  the fractional variability is high in all the observations and almost  independent on the selected energy band. For example in the 7-10 keV energy band the pn light curves are characterized by  a  $F_{\rm var}$ that ranges from $\sim 19$ (OBS3) to $\sim 36$ (OBS1) per cent. We also note that  the variability is present in all the observations regardless of the observed flux, where both the brightest state (OBS1) and the faintest one (OBS2) shows a similar  level of variability.

 \section{Broad band spectral analysis}\label{spectra_std}
We performed all the spectral fit with XSPEC (ver. 12.11) and we included in all the models the      Galactic absorption  in the direction of \sorg\ ($\nhsym=1.9\times 10^{20}$\,\nh, \citealt{nhHI4PICollaboration}),  modelled with the  Tuebingen - Boulder absorption model   (\textsc{tbabs} component in  \textit{XSPEC}, \citealt{Wilms2000}).  We used     $\chi^2$ statistics and   the errors are quoted at the 90 per cent confidence level for one interesting parameter unless otherwise stated. All the parameters are given in the rest frame of \sorg\ ($z=0.03233$, \citealt{Sirressi2019}) and the velocities are all relativistically corrected\footnote{Velocities are    corrected with   the relativistic Doppler formula: $v/c=[(1+z_{\rm o})^2-1]/[(1+z_{\rm o})^2+1]$, where $z_{\rm o}$ is the measured rest-frame blueshift.}.
We fitted the  EPIC-pn and  \nustar-FPMA+FPMB (hereafter FPM) spectra  jointly and considered the 0.3-10 keV and the 3.5-45 keV  data for the EPIC-pn and FPM spectra, respectively. We included a multiplicative constant  between the EPIC-pn and FPM spectra to account for any difference in normalization due to cross-calibration; this was allowed to vary and in all the fits we found that it  ranges between $1.01\pm 0.03$ (OBS2) and $1.12\pm 0.02$ (OBS1). \\ 
 \begin{figure}[t]
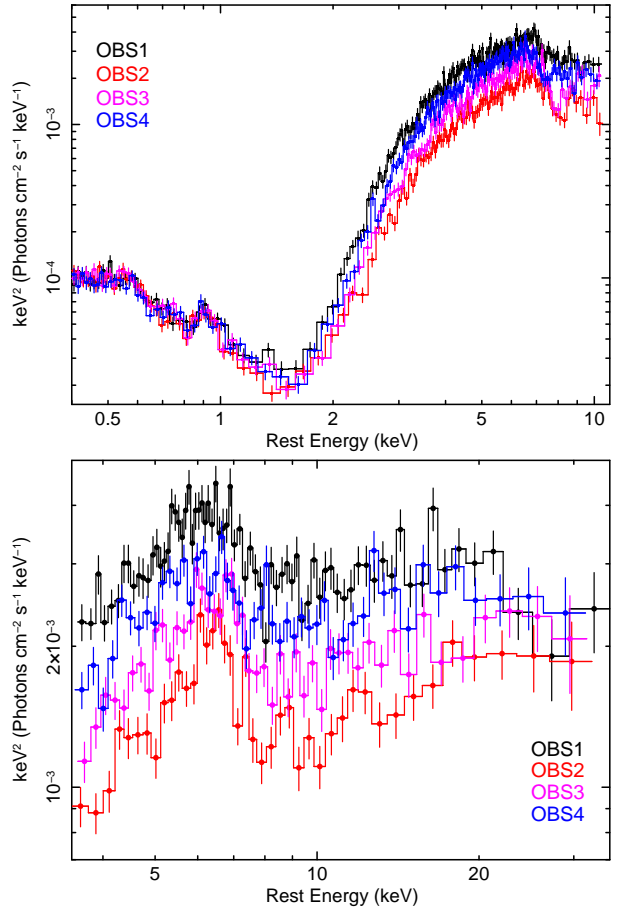
 
\rotatebox{-90}{
\includegraphics[angle=0,width=0.33\textwidth]{fig3a.ps}
\includegraphics[angle=0,width=0.33\textwidth]{fig3b.ps}
}
\caption{{Upper panel: broad-band    (rest frame 0.4-10.5 keV) \xmm\    spectra of the four observations of \sorg\ performed in 2019.  The $E F_{\rm E}$ spectra  are obtained unfolding the data against a simple  power-law model with $\Gamma=2$.  
No variability is present below 2 keV, where the emission is dominated by the scattered component, while clear variability is present at higher energies, where \sorg\ dropped in flux during the OBS2. During OBS3,   \sorg\ was in a  slightly brighter state,  but  with an     absorption structure almost as deep as in OBS2.  Lower panel: \nustar\ spectra of the same observations.   
Like for the \xmm\  data,  the spectra are plotted against a simple $\Gamma=2$ power-law model.  The \nustar\ data independently confirms the presence of the deep absorption structures.}
 \label{fig:compare_eeuf}
 }
\end{figure}

 \subsection{The overall spectral shape}
 In Fig.\ref{fig:compare_eeuf}, we show  the EPIC-pn (upper panel) and \nustar-FPM (lower panel) fluxed spectra of all the 4 observations. The fluxed spectra are obtained 
  by unfolding the data against a power law model with $\Gamma=2$.   As already noted with the previous observations of \sorg\ (B21), the soft X-ray emission as well as the 2-6 keV   spectral curvature, that we can ascribe to the presence of the distant and  neutral or low ionization absorber, do not vary. Conversely, although all the observations show  absorbing structures  at $E>7$\, keV, the depth of these features varies between the observations. In particular they are deeper in OBS2 (red  spectra in Fig.~\ref{fig:compare_eeuf})  and in OBS3 (magenta  spectra in Fig.~\ref{fig:compare_eeuf}) with respect to OBS1 and OBS4, with OBS1 showing the shallower structures. Finally,    as shown in Fig.~\ref{fig:compare_eeuf} (lower panel),  above     15  keV, where the primary continuum emerges, the   \nustar\ data show  that this clearly varies between the observations.    In particular, 
    while during OBS1 and OBS4, \sorg\ was almost at the same intrinsic  flux level, OBS2 and OBS3 are characterized by a lower level of the primary emission. This could be  already anticipated by the inspection of the light curves extracted in the 15-35 keV  energy range, where the mean varied from  $0.051\pm 0.002$ cts s$^{-1}$ in OBS 1 to $0.032\pm 0.001$  cts s$^{-1}$ in OBS 2 (see Table~\ref{fvar}). 
    
\begin{figure}
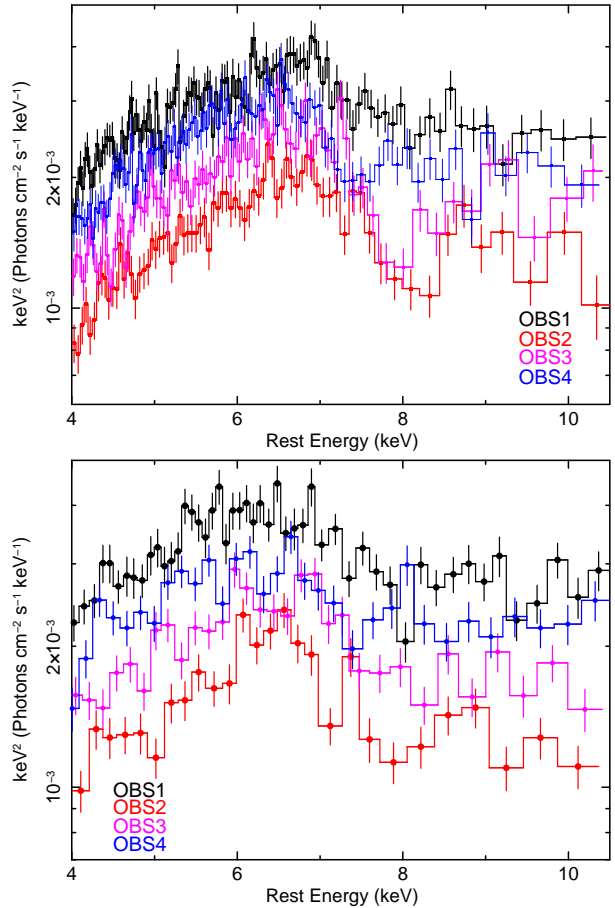
 
\rotatebox{-90}{
\includegraphics[angle=0,width=0.33\textwidth]{fig4a.ps}
\includegraphics[angle=0,width=0.33\textwidth]{fig4b.ps}
}
\caption{{Upper panel: Zoom into the EPIC-pn 4-10.5 keV (rest frame) energy range, showing the different profile of the absorption  structure seen in the 8-10 keV energy range. The $E F_{\rm E}$ spectra  are obtained unfolding the data against a simple  power-law model with $\Gamma=2$.     Lower  panel:  Analogous zoom into the \nustar\ spectra in the 4-10.5 keV energy range. 
The structures are less pronounced during OBS1 and dramatically increase in depth during the OBS2, which was performed  10 days later.  Both OBS2 and OBS3  show two deep absorption troughs at  $\sim 8 $ keV and $\sim 10$ keV. OBS4 caught \sorg\ almost at the same flux level of OBS1, but with a different profile of the absorption structure, with  the low  energy feature being now at $\sim 7.5 $ keV.  
 }
 \label{fig:zoom_eeuf}
 }
\end{figure}

In Fig.~\ref{fig:zoom_eeuf}, we show a zoom into the 4-10.5 keV (rest frame) EPIC-pn (upper panel) and \nustar\ (lower panel) spectra, which nicely show the changes in the absorbing structures. Here, at the beginning of this new observational campaign the absorbing structure is clearly shallower, while a broad and deep absorption trough is present at around 8 keV in the second and third observations (red  and magenta spectra in  the upper and lower panel of Fig.~\ref{fig:zoom_eeuf}) and at around  $\sim 7.5 $ keV in the last one. 

    Notably, all these variations occur on relatively short time-scales, while \sorg\ was brighter in OBS1, 10 days later (OBS2) it becomes clearly fainter and harder, with a possibly   stronger wind. After just over 4 days the intrinsic flux increases and the wind appears to be as strong as in OBS2; then after two more weeks \sorg\ is almost at the same flux level of OBS1. Intriguingly, in OBS4  the absorbing structure appears to be at a lower energy with respect to all the other observations.   

\subsection{The Broadband Baseline Model}
We then proceeded to fit  all the observations with a baseline continuum model and considered all the four observations simultaneously.
We first ignored the energy range where the  Fe-K emission line is expected and where the absorbing troughs are present, namely we ignored  the data in the 6-15 keV energy range. Following the results obtained with the previous observations (see B21), we initially tested a neutral absorber (model \textsc{zphabs} in \textsc{xspec}) to account for the 1.5-6 keV curvature. However, this simple neutral absorber cannot fully reproduce the spectral shape between 1.5-3 keV, leaving some positive residuals, which  suggests that   the absorber could be mildly ionized and thus more transparent at those energies. We thus considered a  fully covering  ionized absorber. This latter  was   modeled with a multiplicative grid of photoionized absorbers generated with the \textsc{xstar} photoionization code (\citealt{xstar}). We adopted a grid that has a low   turbulence velocity ($v_{\rm turb}=200$\,km s$^{-1}$) and covers a wide range of ionization states (log($\,\xi /{\rm erg\,cm \,s^{-1})}$ between $-3$ and  $6$).  
The form of the  baseline model is:
 \begin{equation*}
 \begin{split}
 F(E)=\textsc{tbabs}\times[\textsc{zpow}_{\rm scatt}+{\rm Gauss_{\rm Soft}}+\\
 \,\textsc{mekal}+  \,\textsc{xstar}_{\rm LOW}\times \textsc{zpow}]
\end{split}
\end{equation*}

 where all the emission  components are absorbed by the Galactic absorption (modeled with \textsc{tbabs}; $\nhsym=1.9\times 10^{20}$\,\nh; \citealt{nhHI4PICollaboration}).   For the primary emission (\textsc{zpow}), we allowed both the normalization and the photon index to vary between the observations.  We   allowed the $\nhsym$ of the ionized absorber (\textsc{xstar}$_{\rm LOW}$) to vary, but we assumed that it has the same ionization state in all the four observations. For the soft X-ray emission,  the baseline model includes both the  emission from  a  collisionally ionized plasma (\textsc{mekal} component in \textsc{xspec}, \citealt{Mewe85}; with $kT=0.13\pm 0.01$\,keV) and    an additional Gaussian emission line  (Gauss$_{\rm Soft}$) at $E=0.92\pm 0.01$ keV. As there is no evidence of variability of the soft X-ray emission, we tied all its parameters between the four observation. As noted in  previous works (M19, B21),  when  the  photon index of the scattered component ($\textsc{zpow}_{\rm scatt}$) is allowed to vary,  it  tends to a high value ($\Gamma_{\rm{ soft}}=3.3\pm 0.1$).  
  
  Although simplistic, this model can  reproduce the overall spectral shape ($\chi{^2}/\nu=1276.6/1276$, excluding the 6-15 keV energy range). We found that both the photon indices and normalizations of the primary power-law component ($N_{\rm PL}$) vary across the monitoring campaign. For example, in OBS1 we  measure $\Gamma=2.55\pm 0.05$  and $N_{\rm PL}= (1.3\pm 0.1) \times 10^{-2}$\,ph cm$^{-2}$ s$^{-1}$ keV$^{-1}$, while in  the harder and faint state (OBS2) we derive $\Gamma=2.22\pm 0.06$  and $N_{\rm PL}= (3.8\pm 0.4) \times 10^{-3}$\,ph cm$^{-2}$ s$^{-1}$ keV$^{-1}$.   Conversely,  as seen in the previous observations the absorber responsible for the 3-6 keV curvature does not vary between the observations with an average column density of $\nhsym\sim 8.3\times 10^{22}$\, \nh.       The ionization state is    log($\,\xi /{\rm erg\,cm \,s^{-1})}=0.31\pm 0.02$.\\

\begin{figure}
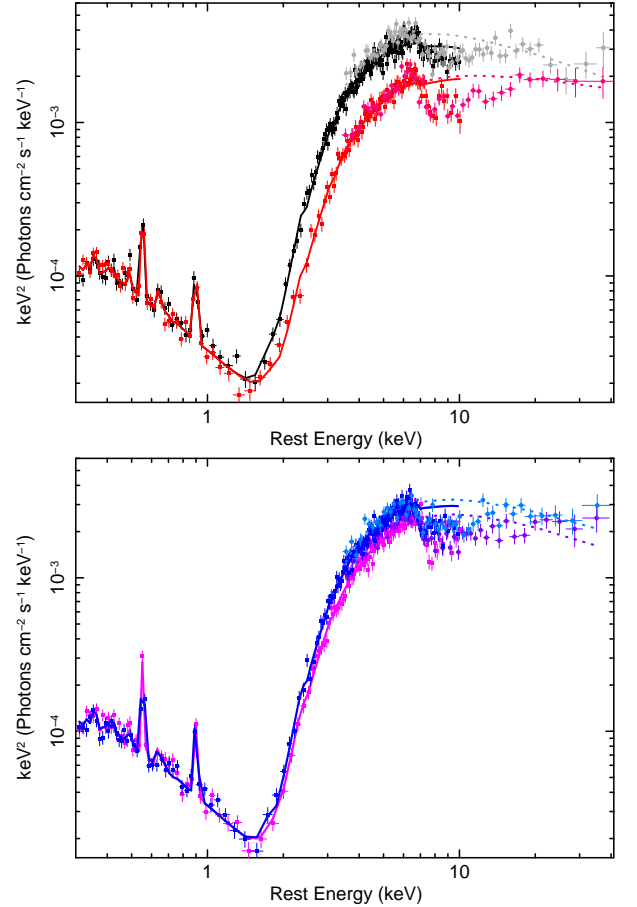
 
\rotatebox{-90}{
\includegraphics[angle=0,width=0.33\textwidth]{baseline_1v2_restnew.ps}
\includegraphics[angle=0,width=0.33\textwidth]{baseline_3v4_restnew.ps}
}
\caption{{Broad band (rest frame) pn and FPMA spectra unfolded against the baseline continuum model described in \S  4.2. Upper panel:  OBS1 (pn and FPM are shown in black squares and grey filled circles, respectively) and OBS2 (pn and FPM are shown in red  squares and  dark red filled circles, respectively). Lower panel: OBS3 (magenta squares and purple filled circles ) and OBS4 (blue squares and  light blue filled circles). Here the baseline continuum model is able to account for the overall spectral shape of the X-ray emission of \sorg. 
 }
 \label{fig:baseline_eeuf}
 }
\end{figure}

\begin{figure}
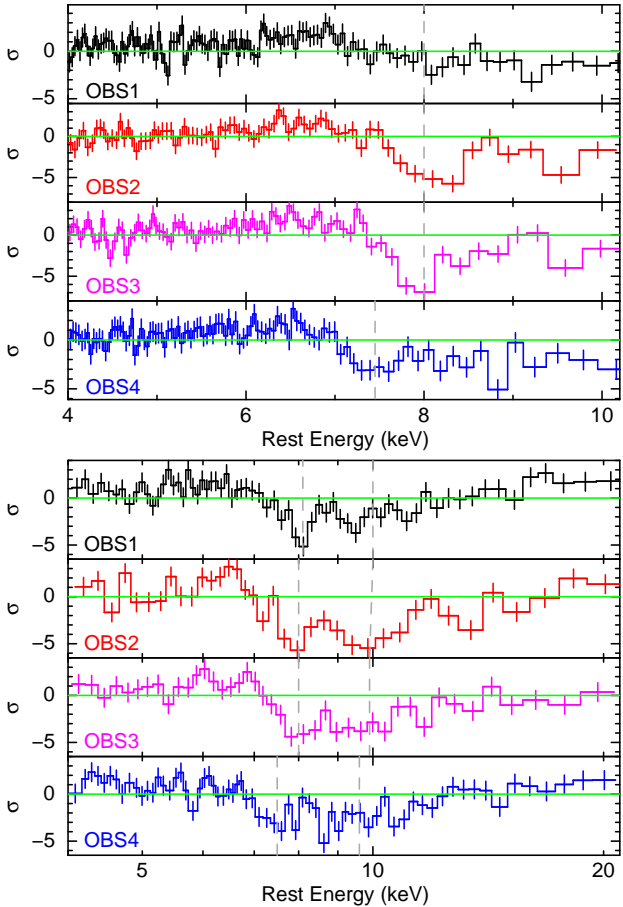
 
\rotatebox{-90}{
\includegraphics[angle=0,width=0.33\textwidth]{fig6a.ps} 
\includegraphics[angle=0,width=0.33\textwidth]{fig6b.ps} 
}
\caption{{ Zoom into the  residuals (as data-model/error), against the baseline continuum model showing the different profiles of the absorption  structures seen in the 7.5-10 keV energy range.  OBS1, OBS2, OBS3 and OBS 4 are shown in black, red,  magenta and blue  respectively. In the upper panel we report the  EPIC-pn 4-10 keV residuals, while  the \nustar\ 4-20 keV residuals are shown in the lower panel. Absorption structures (marked with the vertical dashed lines) are present in all the observations, but they show a remarkable variability in depth and energy, whereas they are less  pronounced during OBS1 and increase in depth during OBS2.    A possible second structure is also visible in all the residuals at around 10 keV. This could be either associated to a  higher-velocity wind phase or ascribed to  a contribution from higher-order Fe K absorption (see \S  4.3 and 5.1). 
}
 \label{fig:del_baseline}
}
\end{figure}

In Figure~\ref{fig:baseline_eeuf}, we show the resulting    pn and FPM spectra unfolded against this baseline model where we included  the data inbetween 6 and 15 keV.   As shown in  Figure~\ref{fig:del_baseline}, where we report the residuals (plotted in terms of data - model / error) to this initial baseline continuum model, a clear   deficit   of counts is present above 7 keV in all of them. In particular, a  strong Fe-K absorption structure is clearly visible at $\sim 8$ keV in the first three observations and at $\sim 7.5$ keV in the last observation. If associated with \fexxvi\  ($E_{\rm Lab}=6.97$ keV) the inferred outflow velocity is of the order of $\sim -0.14\,c$ in OBS1, OBS2 and OBS3, while in OBS4 the inferred velocity is lower and of the order of $\sim -0.07\, c$. 
 The depth  and breath of the troughs   appear  to be variable, where the absorption is deeper in OBS2 and OBS3 and shallower  in OBS1. 
 A second  structure is clearly visible in all the residuals around $\sim 10-11$\,keV and    the structure is again more pronounced in OBS2 and OBS3.
A broad  excess is visible  in the 6-7 keV energy range suggesting the presence of a possibly broad Fe-K emission line,  while there is no evidence for a narrow \feka\ emission line, which could originate from reflection off the putative pc-scale torus. Similarly, no clear  excess,  that could be ascribed to a strong Compton reflection, emerges above 15 keV.

  \subsection{Photoionisation Modelling of the Fe K Absorption}
  We then proceeded to fit  the broad band data including both a  Gaussian emission line, to account for the possibly  broad Fe-K emission line, and  the ionized absorber.  
   In order to determine the optimal multiplicative grid of photoionized absorbers,  we first note that   the underlying continuum is similar to what we measured  in the past observations, where the soft (between 0.3 and 1.2 keV) photon index is $ \Gamma \sim 3$   and the hard X-ray $\Gamma$   is between 2 and 2.6. We therefore
 adopted an absorption   grid similar to the one used by B21\footnote{To generate the grid we used   a SED similar to IZW1, where the soft (between 0.3 and 1.2 keV) photon index ($\Gamma$) is $\sim 3$   and the hard X-ray $\Gamma$  is 2.2 (see Reeves \& Braito 2019)}, which was generated with \textsc{xstar} and  optimised for high column density  ($\nhsym=10^{22}$  - $2\times10^{24}$ \nh) and high ionisation (log($\,\xi /{\rm erg\,cm \,s^{-1})}= 3$ - $7$) absorbers.

 We then inspected the main absorption structures seen   between  $\sim 8-12$ keV, to determine the optimal velocity broadening. To this end, we considered OBS2 and OBS3,   where the  absorption structures are deeper and better defined. We thus   added two Gaussian absorption lines, we allowed their energy centroids and normalizations to vary between the two observations, but assumed that they have the same width. The first absorption line   accounts for the low energy trough; its  energy centroid is found to be around 8 keV   ($E=8.08 \pm 0.09$ keV and $E=8.07 \pm 0.10$\, keV    for   OBS2 and OBS3, respectively)  with an $EW$ of   $-475 \pm 80$ eV in OBS2  and $-472\pm 75$ eV in OBS3. A second Gaussian absorption line, which     accounts for the residuals at $\sim 10-11$\, keV, is  detected  at $E=10.0\pm 0.2$ keV in OBS2   ($EW=-550\errUD{90}{110}$\,eV)
 and $E=9.9\pm 0.2$ keV in OBS3 ($EW=-410\pm 90$\,eV).  We note that both  absorption lines are highly   significant with   $\Delta \chi^2/\nu=208/5$ and $\Delta \chi^2/\nu=152/4$ for the lower and higher energy feature, respectively. If compared to the expected energy of H-like Fe\,\textsc{xxvi} Ly$\alpha$ line ($E_{\rm Lab}=6.97$\,keV), the measured energies  of these two absorption lines would correspond to  outflow  velocities of  $v/c\sim -0.15 $ and $v/c\sim-0.34$,  respectively. The profiles are, as   expected from the inspection of the residuals,   broadened  with a best-fit   width of $\sigma=510^{+70}_{-60}$\,eV,  corresponding to a velocity broadening of $\sigma_{\rm v}\sim20000$\,km\,s$^{-1}$.  Taking into account the measured broadening of the troughs, we chose a grid generated assuming a high turbulence velocity  ($v_{\rm turb}=25000$\,km s$^{-1}$), we however tested lower turbulence velocities and found that they generally resulted in a statistically worse fit. \\
 
 \begin{deluxetable*}{llcccc}[t]
 \tablecaption{\label{tab:xstar}Photoionization Modeling of the Wind. }
\tablewidth{0pt}
\tablehead{
\colhead{Model Component } & \colhead{Parameter} & \colhead{OBS1}   & \colhead{OBS2}  & \colhead{OBS3} & \colhead{OBS4}} 
\startdata
\multicolumn{6}{c}{Best Fit Parameters for the    baseline continuum$^{a}$.}\\
\hline\\
Primary PL& $\Gamma$ & 2.57$\errUD{0.04}{0.03}$     & 2.30$\errUD{0.07}{0.04}$   & 2.37$\errUD{0.05}{0.03}$   &2.44$\errUD{0.05}{0.03}$    \\
 & $N_{\rm PL}{^b}$& 	      $14.0\errUD{1.2}{0.79}$     & $4.6\errUD{0.9}{0.3}$ &$6.7\errUD{0.9}{0.5}$ & $9.4\errUD{0.9}{0.5}$ \\
 \hline\\
  Absorber & $\nhsym$ ($\times 10^{22}$\,\nh) &     $8.35\errUD{0.28}{0.03}$    & $8.35^t$ &$8.35^t$  & $8.35^t$  \\
  & log$\xi$& 	      $-0.29\errUD{0.01}{0.02}$  & $-0.29^t$ & $-0.29^t$   & $-0.29^t$ \\
 \hline\\
\multicolumn{6}{c}{Best Fit Parameters for the    variable $\nhsym$ case.  The   statistics for the model is $\chi^2/\nu=1868.5/1784$.}\\
\hline\\
Zone 1 & $\nhsym$($\times 10^{23}$\nh) &$4.2\errUD{1.0}{0.7}$    & $7.4\errUD{1.6}{1.0}$  & $7.2\errUD{1.4}{0.9}$  & $5.3\errUD{1.4}{0.9}$       \\
 & log$\xi$& 	      $4.90\errUD{0.07}{0.08}$  & $4.90^t$ &$4.90^t$   & $4.90^t$ \\
& $v_{\rm out1}/c$& 	      $-0.154\pm0.015$  & $-0.153\pm0.012$&$-0.157\pm0.011$  & $-0.077 \pm0.013$ \\
\hline \\
Zone 2 & $\nhsym$($\times 10^{23}$\nh) &$3.8\errUD{1.0}{0.9}$    & $7.1\errUD{1.6}{1.5}$  & $4.7\errUD{1.5}{1.1}$  & $5.3\errUD{1.4}{1.0}$       \\
 & log$\xi$& 	      $4.90^t$  & $4.90^t$ &$4.90^t$   & $4.90^t$ \\
& $v_{\rm out1}/c$& 	      $-0.33\pm0.02$  & $-0.36\pm0.02$&$-0.34\pm0.02$  & $-0.25 \pm0.02$ \\
 \hline  \\
\multicolumn{6}{c}{Best Fit Parameters for the    variable ionization case. The   statistics for the model is $\chi^2/\nu=1870.7/1787$.}\\
 \hline\\
Zone 1 & $\nhsym$($\times 10^{23}$\nh) &$9.4\errUD{1.4}{1.2}$    & $9.4^t$ & $9.4^t$   & $9.4^t$        \\
 & log$\xi$& 	      $5.32\errUD{0.11}{0.08}$  &  $5.02\errUD{0.09}{0.09}$ & $5.10\errUD{0.09}{0.08}$  &  $5.19\errUD{0.09}{0.08}$ \\
& $v_{\rm out1}/c$& 	      $-0.145\pm0.015$  & $-0.151\pm0.012$&$-0.149\pm0.011$  & $-0.072 \pm0.010$ \\
 \hline  \\
Zone 2 & $\nhsym$($\times 10^{23}$\nh) &$8.2\errUD{1.2}{0.9}$    &  $8.2^t$  & $8.2^t$  & $8.2^t$       \\
 & log$\xi$& 	      $5.32^t$  & $5.02^t$ &$5.10^t$   & $5.19^t$ \\
& $v_{\rm out1}/c$& 	      $-0.32\pm0.02$  & $-0.36\pm0.02$&$-0.34\pm0.02$  & $-0.27 \pm0.02$ \\
\hline  \\
\multicolumn{2}{l}{$F_{(2-10)\,\mathrm {keV}}$ ($\times 10^{-12}$\,\flux) }  & 4.9  & 2.5  & 3.2  & 4.0       \\
\multicolumn{2}{l}{$F_{(10-40)\,\mathrm {keV}}$ ($\times 10^{-12}$\,\flux) }  & 5.5   &  3.5  &   4.3& 5.2  \\
\multicolumn{2}{l}{$L_{(2-10)\,\mathrm {keV}}$ ($\times 10^{43}$\,\lum) }  & 3.7   & 1.8  &   2.3& 3.0  \\
\enddata
 \tablenotetext{a}{The baseline continuum model parameters are not affected by the assumption of a variable column density or  ionization of the  wind components.}
\tablenotetext{b}{The normalizations are in  units of $\times 10^{-3}$ ph cm$^{-2}$ s$^{-1}$ keV$^{-1}$ at 1 keV.}
\tablenotetext{t}{Denotes that the parameter was tied}
\end{deluxetable*}

In the previous observations of \sorg, two  multiplicative grids of  ionised absorption models, were required to account for all the absorption structures that were detected (see B21 and reference therein).  Likewise, in this new set of observations a single photoionized outflowing absorber  cannot account for both the absorbing structures that are  visible in the residuals. This is  mainly due to the depth  of the higher energy feature. Precisely, if we associate the absorption line detected at $\sim 8.1$\, keV   with  \fexxvi\ Ly$\alpha$ ($E_{\rm Lab}=6.97$ keV) blue-shifted by $v=0.15\,c$, the corresponding \fexxvi\ Ly$\beta$ ($E_{\rm Lab}=8.25$ keV) would be blue-shifted  to $\sim 9.5$\,keV, which is almost consistent with the energy of higher energy  feature. However, this line is expected to be weaker than the  \fexxvi\ Ly$\alpha$ and not of a similar  EW. However, this  is a limitation of the modelling with  a   grid of photoionized absorber generated with \textsc{xstar}, where the  emission of the wind is not self-consistently accounted for and crucially the breadth of the line is purely ascribed to turbulence (see below). \\

 We thus  applied a model defined as: \begin{equation*}
   \begin{split}
 F(E)=\textsc{tbabs}\times[\textsc{zpow}_{\rm scatt}+ 2 \,\textsc{Gaus}_{\rm Soft}+\,\textsc{mekal}+\\ 
 {\rm \feka} +\,\textsc{xstar}_{\rm LOW}\times \textsc{xstar}_{\rm FeK,1}\times\textsc{xstar}_{\rm FeK,2}\times\textsc{zpow}]
\end{split}
\end{equation*}

Here, we added a Gaussian emission line to account for the \feka\ emission line,   two    highly ionized  outflowing absorber   (\textsc{xstar}$_{\rm FeK,1}$ and \textsc{xstar}$_{\rm FeK,2}$; hereafter zone1 and zone2) and replaced the neutral absorber  with a  mildly ionized absorber (\textsc{xstar}$_{\rm LOW}$). We then proceed to perform a joint fit for all the observations. \\
As there is no evidence of variability of the soft X-ray emission, we again tied all the parameters of the soft X-ray components. We also tied the parameters of the mildly ionized absorber (\textsc{xstar}$_{\rm LOW}$) and of  the \feka\ emission line. The normalization and photon index of the primary emission were allowed to vary. 

In order to quantify the apparent disk wind variability    and taking into account   the known degeneracy between the ionization and $\nhsym$,  
 we first allowed the column densities of the two highly ionized absorbers  (zone1 and zone2) to vary,  while the wind ionization was assumed to remain constant and to  be the same for both the zones. We also allowed the velocities of both zone1 and zone2 to vary between the observations. In Table~\ref{tab:xstar}, we list the results of this fit; overall the fit statistic is good ($\chi^2/\nu=1868.5/1784$  and  only weak negative residuals are present in the 10-15 keV energy range ($\chi^2/\nu=421.1/316$ in the 9-30 keV range). We note that both the zones are statistically required and removing either one of them results in much worse fits,  where we have $\Delta \chi^2/ \Delta \nu=364.7/9$ for zone1    and  $\Delta \chi^2/\Delta \nu=203.2/8$ for zone2.

This model confirms that \sorg\ was caught in the intrinsically brightest  state during OBS1 and in the faintest state in OBS2.  During  OBS3, \sorg\  starts to get brighter and in the last observation is  almost at the same flux level of OBS1 (see Table~\ref{tab:xstar}).  As anticipated by  the inspection of the residuals, OBS1 is characterized not only by a higher intrinsic flux   but also by weaker wind with   $N_{\rm H\, OBS1,1}= 4.2\errUD{1.0}{0.7}\times 10^{23}$\,\nh\ and $N_{\rm H\,OBS1,2}= 3.8\errUD{1.0}{0.9}\times 10^{23}$\,\nh\ for the lower (zone1) and higher (zone2) velocity component of the wind. Conversely, OBS2    not only caught \sorg\ at an intrinsically low flux state, but  with also  a higher density wind, where the column density of both zones increases to  $N_{\rm H\, OBS2,1}= 7.4\errUD{1.6}{1.0}\times 10^{23}$\,\nh\ and $N_{\rm H\, OBS2,2}= 7.1\errUD{1.6}{1.5}\times 10^{23}$\,\nh, respectively.  
The outflowing velocities of the wind are similar in the first three observations and they are of the order of $v_1/c\sim - 0.15$ and $v_2/c\sim - 0.33$, for the slow and the possibly fast component. The exception is OBS4, where  both the velocities are lower  ($v_1/c=-0.077\pm0.013 $ and $v_2/c=-0.25\pm 0.02$) than the corresponding ones measured in the previous observations (see Table~\ref{tab:xstar}).\\

 Alternatively, the change in the opacity of the wind can be parametrized by a decrease of the ionization in response to a lower intrinsic flux. 
 In order to test this scenario, we assumed that the column density of each zone  remains constant, while the ionization was allowed to vary in each of the observations; we still assumed that the two zones have the same ionization as each other. As in the previous test, we allowed the wind velocities to vary. From a statistical point of view, this model is undistinguishable from the previous one with $\chi^2/\nu=1870.7/1787$ ($\chi^2/\nu=423.1/319$ in the 9-30 keV range), and also in this case there are no strong residuals.  
 The increase of the opacity of the wind in OBS2 is now explained with a decrease of the ionization from log$\,\xi=5.32\errUD{0.11}{0.08}$ (OBS1) to log$\,\xi=5.02\errUD{0.09}{0.09}$ (see Table~\ref{tab:xstar}). In Fig.~\ref{fig:xstar_fit}, we show a comparison   of  Fe-K profiles, for the variable ionization case,  in the two most different states observed during  the campaign (OBS2 and OBS4). We can see that the deeper absorption features seen in OBS2 can be well reproduced by a decrease of the wind ionization. It  is also noticeable that the profile of the absorption troughs  is less blue-shifted in OBS4 than in OBS2. We note that the ionization varies in proportion with the changes of the 2-10 keV luminosity, which suggest that the flow could be  in photo-ionization equilibrium  and it  recombines in  response to a lower luminosity of the X-ray source. Interestingly, once again we find that, while  the outflowing velocities  remained   unchanged, within the errors, during the first three observations, they differ in OBS4. Similarly to the variable $\nhsym$ model, the wind in OBS4 is slower  (see Table~\ref{tab:xstar}).   We note that in both  scenarios (variable $\nhsym$ or variable ionization) the wind is highly ionized, such as the observed  absorption is basically always due to \fexxvi.

  \section{The Disk Wind  Model}\label{spectra_diskw}

We then proceeded to model the variable disk wind in  \sorg\ with a self consistent  disk wind model. To this end we used a table  of synthetic wind spectra,    
that were generated using the radiative transfer diskwind code developed by \citet{Sim2008, Sim2010a,Sim2010b}. The spectral tables were computed  for smooth, steady-state 2.5D\footnote{We refer as 2.5D because, while the velocity field is 3D, all the other parameters (e.g. density, ionisation) of the wind are axisymmetric.} bi-conical winds.     This  disk wind model treats self-consistently the absorption and re-emission from the wind and  parameterizes  the velocity field through the wind. Thus  the  spectra contain both the emission transmitted through the wind as well as the reflection or scattered emission from the wind, the latter includes also the Fe-K emission. The model also computes the ionization structure through the wind; we note that the model  contains  extensive atomic data and covers   a wide range in ionization (for example from Fe\,\textsc{x} to \fexxvi). The calculation is not limited to iron but it  includes  lighter elements as well.  The wind geometry is of a bi-conical wind  that is launched from an inner radius $R_{\rm min}$ and an  outer radius $R_{\rm max}$; the wind opening angle is set by a geometrical parameter $d$, which is defined as the distance from the focus point of the wind    below the origin (in units of $R_{\rm min}$). This parameter also determines the inclination of the wind with respect to the equatorial plane. 
A schematic view of the inner disk wind model geometry was presented by \citet{Sim2008} and it is also reported by \citet{ReevesBraito2019} (see their Fig.~8). \\ As discussed in   previous works, which presented and tested this disk wind model  (\citealt{Sim2008,Sim2010a,Tatum2012,Reeves2014,ReevesBraito2019}),  an important parameter  is the orientation of the observer line of sight with respect to  the wind. This is defined  through the parameter $\mu=cos\theta$, where 
  $\theta$ is the angle between the polar axis of the wind and the observer's line of sight. At low inclination angles the observer can view directly the primary X-ray emission without intercepting the wind, therefore the spectra have little or no absorption. However,  the computed  spectra have the contribution from X-ray reflection  off the surface of the wind, which includes also a broadened \feka\ emission line. At higher values of $\theta$ (lower values of  $\mu$), the observer's line of sight intercept more of the wind; the computed spectra  contain blue-shifted absorption features and also the  scattered emission.  \\
  
  In order to model the X-ray spectra of \sorg, we generated a grid of spectra assuming   an inner launch radius of $R_{\rm min}= 64 R_g$, where $R_g$ is the gravitational radius defined as $R_g= GM/c^2$ and $R_{\rm max}=1.5R_{\rm min}$. Thus $R_{\rm min}$ corresponds to    the escape radius for a wind with a terminal velocity of $v_\infty =-0.177 c$. When generating the grid of spectra, we adopted $d=1$, which at large radii corresponds to a wind with an opening angle of $\pm  45 ^\circ $ with respect to the polar axis of the wind.  
The model also assumes that the X-ray source is concentrated in a region of $6R_g$ centered at the origin. We note that this disk wind model takes into account  special relativistic effects. \\
  
The  synthetic wind spectra are then  tabulated in a  multiplicative table that can be loaded within \textsc{xspec}. The main parameters that define the wind condition are: 
\begin{itemize}
\item   The slope of the input continuum. The continuum is set as power-law with $\Gamma$ ranging from 1.6 to 2.4 with a linear step of 0.2, calculated over the 0.1-500 keV energy range.  
\item Terminal velocity. The terminal velocities  are set by the choice of the inner wind radius ($R_{\rm min}$) and the   terminal velocity parameter $f_{\rm v}$. This latter parameter determines  the terminal velocity of the wind streamline from the escape velocity  at its base  with $v_\infty=f_{\rm v} \sqrt{2GM_{\rm BH}/R}$. The terminal velocity is then derived by varying the parameter $f_{\rm v}$, for the assumed launch radius of $R_{\rm min}= 64 R_g$. The adopted table was generated for 8 velocities with $f_{\rm v}$ ranging from 0.25 to 2.0 (in linear increments). These $f_{\rm v}$ values correspond to terminal velocities of $0.0445c-0.356c$.
\item Inclination angle. This is defined with respect to the polar axis of the wind ($\theta$) and 
 is parametrized via $\mu=cos\theta$. Here $\mu$ covers the 0.025-0.975 range in 20 steps (with $\Delta\mu=0.05$). Note that, having set the opening angle to $45^\circ$ for $\mu>0.7$ (i.e. $\theta <45^\circ$), the observer's line of sight does not intercept the wind. 
\item Mass outflow rate. It is expressed in Eddington units as $\dot M=\dot M_{\rm out}/\dot M_{\rm Edd}$ and it covers the 0.02-0.68 range in 12 logarithmic steps.
Clearly, higher  $\dot M$ give spectra with  stronger absorption and emission features.
  \item Ionizing X-ray luminosity ($L_{\rm X}$).  It is defined as the percentage of the 2-10 keV luminosity  with respect to the Eddington luminosity ($L_{\rm X}=L_{\rm 2-10\, keV}/L_{\rm Edd}$). Like the mass outflow rate it is defined in Eddington units, which makes   both   parameters invariant with respect to the black hole mass. The  X-ray luminosity sets the ionization state of the flow, where higher values of $L_{\rm X}$ result in more ionized winds.  $L_{\rm X}$ covers the 0.025-2.5\% of $L_{\rm Edd}$  in 9  equally spaced logarithmic steps.

\end{itemize}

The disk-wind parameter space has been extended from previous work (e. g. \citealt{Reeves2014,ReevesBraito2019}) and hence the table grid is now composed of 86400 synthetic spectra with the free parameters described above ($\Gamma$, $f_{\rm v}$, $\mu$, $\dot M$ and $L_{\rm X}$, Matzeu et al. in prep.). We note that, when fitting within \textsc{xspec}, the best fit parameters and errors are determined  through interpolation,  if they fall in between two grid points.

   \begin{deluxetable*}{llcccc}[t]
\tablecaption{Disk wind model.   The   statistics for the model is $\chi^2/\nu=1881.8/1792$ }
\tablewidth{0pt}
\tablehead{
\colhead{Model Component } & \colhead{Parameter} & \colhead{OBS1}   & \colhead{OBS2}  & \colhead{OBS3} & \colhead{OBS4}} 
\startdata
\multicolumn{6}{c}{Best Fit Parameters for the    baseline continuum$^{a}$}\\
\hline\\
Primary Power-law & $\Gamma$ & $2.37\pm0.04$   & $2.14\pm0.04$   & $2.21\pm0.04$ &$2.28\pm0.04$    \\
 & $N_{\rm PL}{^a}$& 	      $1.65\errUD{0.12}{0.12}$     & $0.66\errUD{0.06}{0.06}$ &$0.98\errUD{0.08}{0.08}$ & $1.41\errUD{0.13}{0.12}$ \\
 \\
 \hline\\
& $\dot{M}_{\rm out}/\dot{M}_{\rm Edd}$$^{b}$ &$0.40\errUD{0.06}{0.05}$    & $0.51\errUD{0.05}{0.05}$  & $0.50\errUD{0.06}{0.06}$  & $0.19\errUD{0.03}{0.03}$       \\
Disk wind parameters$^{c}$  &  $L_{2-10}/L_{\rm Edd} ^b$  & $0.63\errUD{0.15}{0.10}$  & $0.30^t$ &$0.39^t$   & $0.51^t$ \\
  & $\mu=\cos\theta$$^d$& 	$0.51\errUD{0.02}{0.01}$  & $0.51^t$ &$0.51^t$   & $0.51^t$ \\
 & $v_{\infty}/c ^e$ & 	      $-0.205\errUD{0.009}{0.009}$  & $-0.210\errUD{0.010}{0.009}$&$-0.194\errUD{0.009}{0.009}$  & $-0.074 \errUD{0.014}{0.010}$ \\
\\
\hline  \\
\multicolumn{2}{l}{$F_{(2-10)\,\mathrm {keV}}$ ($\times 10^{-12}$\,\flux) }  & 4.9  & 2.5  & 3.2  & 4.0       \\
\multicolumn{2}{l}{$F_{(10-40)\,\mathrm {keV}}$ ($\times 10^{-12}$\,\flux) }  & 6.0   &  3.7  &   4.5& 5.3  \\
\multicolumn{2}{l}{$L_{(2-10)\,\mathrm {keV}}$ ($\times 10^{43}$\,\lum) }  & 5.6   & 3.1  &   4.2& 5.4  \\
\enddata
 \tablenotetext{a}{The normalizations are in  units of $\times 10^{-2}$ ph cm$^{-2}$ s$^{-1}$ keV$^{-1}$ at 1 keV.}
  \tablenotetext{b}{Mass outflow rates in   Eddington units.}
 \tablenotetext{c}{Percentage of the  (2 - 10 keV) ionizing luminosity to Eddington luminosity; note that for OBS2, OBS3, OBS4 the parameter was tied according to the  ratios of the 2-10 keV fluxes.}
 \tablenotetext{d}{Cosine of wind inclination, with respect to the Polar axis.}
 \tablenotetext{t}{Denotes that the parameter was tied}

 \label{tab:diskwind}
\end{deluxetable*}

 \subsection{Application of The Disk Wind Model}
 We then applied the disk wind model to the four observations, replacing the two  \textsc{xstar} grids  and the \feka\ emission line with the \textsc{diskwind} table. As per the photoionization modelling, we tied all the parameters of the soft X-ray emission and of the low ionization absorber. For the diskwind component,  we tied  the inclination angle as it is unlikely  that the inclination of the wind changes dramatically between the observations, especially when we consider that they are relatively close in time. The photon-index, the  mass outflow rate, the terminal velocity ($f_{\rm v}$) and the normalization of the primary power-law continuum were all allowed to vary. Regarding the ionizing X-ray luminosity    ($L_{\rm X}$), it  was allowed to vary for the first observation, while for the remaining three  observations, we assumed it rescales   with the  normalization of the primary power-law. 
 
 In Table~\ref{tab:diskwind}, we summarize the results of the diskwind fit, we note that the model is able to reproduce the data and it returns a fit statistic of $\chi^2/\nu=1881.8/1792$.   If we  consider the energy range where the wind imparts its structures (7-30 keV)  the fit  clearly improves returning a $\chi^2/\nu=411.3/324$,   while the  \textsc{xstar} model resulted in a  $\chi^2/\nu=423.9/319$ (variable ionization case).  Even more striking is the improvement above 9  keV, here the fit with the \textsc{xstar} model  returned, in the 9-30 keV range, a   $ \chi^2/\nu=206.0/133$ (variable ionization case), while the  \textsc{diskwind} model returns  $\chi^2/\nu=170.2/138$. This is mainly explained by the better modelling of the continuum curvature between 9-30 keV (see below).
  
\begin{figure} 
\rotatebox{-90}{
\includegraphics[angle=0,width=0.33\textwidth]{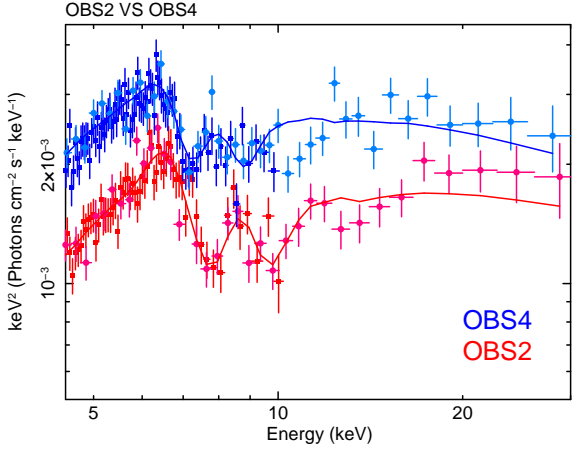} 
}
\caption{{OBS2 and  OBS4  spectra fitted with the best fit  \textsc{xstar} model for a variable ionization.  The {\it XMM}-pn and \nustar\ data are shown with filled square and   circle data points, respectively. For clarity we show a zoom into the 4.5-30 keV energy range.   OBS2 and OBS4 are representative of the two most different states observed during the monitoring campaign, where  the absorption feature is at its deepest during OBS2 and slower during OBS4.  The  deeper absorption troughs  seen in OBS2, can be accounted for by decrease in ionization (from log$\xi \sim 5.19$ to log$\xi \sim 5.02$ ). Alternatively the change in opacity can also be modelled by an increase of the $\nhsym$ (for a constant ionization) by a $\Delta \nhsym \sim 2\times 10^{23}$\,\nh.     
}
 \label{fig:xstar_fit}
}
\end{figure}

\begin{figure}[t] 
\rotatebox{-90}{
\includegraphics[angle=0,width=0.33\textwidth]{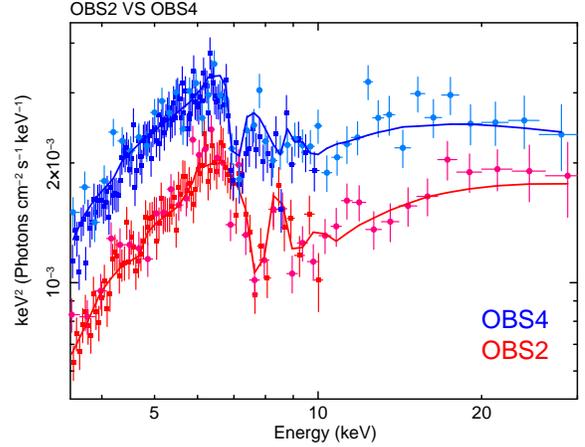} 
}
\caption{{OBS2 and  OBS4  spectra fitted with the disk wind model.  The {\it XMM}-pn and \nustar\ data are shown with filled square and   circle data points, respectively. The model can now reproduce  all the Fe K profile with a single zone.  The  wind terminal velocity is of  $v_{\infty,2}/c=-0.21\pm 0.01$ in OBS2 and $v_{\infty,4}/c=-0.074\errUD{0.014}{0.010}$ in OBS4. Note that the disk wind model includes both the emission   transmitted through the wind and reflected/scattered  from the wind, including the  Fe-K emission. It is noticeable how the model can now better reproduce the  9-30 keV data, when compared to the \textsc{xstar} model as shown in Fig.~\ref{fig:xstar_fit}. 
}
 \label{fig:diskwind_fit}
}
\end{figure}

 We determined $\mu=0.51\errUD{0.02}{0.01}$, corresponding to an inclination angle of $\sim 60^\circ$; thus our line of sight goes through the outflow  and intercepts the inner streamlines of the wind.  The terminal velocities  that we derived with the diskwind model confirm that during OBS4 the wind appears to be slower with respect to the first three observations; the terminal velocity measured in OBS4 is $v_{\infty,4}=-0.074 \errUD{0.014}{0.010} c$, while in OBS1, OBS2 and OBS3 is of the order of  $v_\infty\sim -0.2c$ (see Table~\ref{tab:diskwind}). Thus the  difference in the measured outflow velocities does not depend on the model assumed to fit the absorption structures.  In Fig.~\ref{fig:diskwind_fit}, we show  the      diskwind fit to OBS2  and OBS4. Two main things emerge   with respect to the analogous   Fig.~\ref{fig:xstar_fit}; first of all we note that the diskwind model is able to account for the emission as well as the absorption at  Fe-K, secondly  a single diskwind component is now able to account for the higher energy  absorption structures. Several aspects of the disk wind model can explain this  result.   The main one is that the breadth of the absorption features is not simply due to a micro-turbolence velocity, but it reflects the  range of velocities that our line of sight intercepts.  A second important factor is that the diskwind model self-consistently includes the  emission from the wind,  both the  scattered emission    and  the Fe-K emission lines.  Here, the underlying continuum at $E\sim 10$ keV  is different with respect to the \textsc{xstar}-grid model, because it incorporates the  emission reflected of the surface of the disk wind, which adds some extra curvature between 10-30 keV. Furthermore,   both the \fexxvi-Ly$\alpha$ and  \fexxvi-Ly$\beta$  absorption features are accompanied by the corresponding emission lines.  In other words, the  apparent high EW of the  higher energy absorption feature, which drives the requirement of a second outflowing absorber with the \textsc{xstar} model,  is artificially increased by  the higher level of the intrinsic continuum.  The difference in the continuum, can be also seen by comparing the overall best-fit obtained with the  \textsc{xstar}   (see Fig.~\ref{fig:xstar_fit})  versus  and the  diskwind model (see Fig.~\ref{fig:diskwind_fit}). It is clear that the diskwind model can better reproduce the curvature in the 9-30 keV range, where the \textsc{xstar} model marginally overpredicts the  observed X-ray emission. This  explains the  improvement in the fit statistic over the 9-30 keV range.  \\ 

 For OBS1,  the derived ionizing X-ray luminosity is  $L_{\rm X}=0.63\errUD{0.15}{0.10}$\% of $L_{\rm Edd}$  (for OBS1), which for a black hole mass of $\sim 10^8 M_\odot$ corresponds to a $L_{\rm 2-10 keV}\sim 6\times 10^{43}$\,\lum (see Table~\ref{tab:diskwind}). We note that this is similar to the measured 2-10 keV luminosity, which is of the order of $\sim 5.6\times 10^{43}$\,\lum. This implies that wind ionization is close to the expectation  given the observed X-ray luminosity.  The best-fit  with the diskwind model suggests that during the  first three observations  the overall properties of the disk wind of \sorg\ are rather similar. The variable wind opacity here is determined by the changes of the illuminating X-ray luminosity, which sets the wind ionization.  Here the  wind terminal velocities are all of $\sim -0.2 c$ and the  mass outflow rate is only  marginally  higher during OBS2 and OBS3, where it is of the order of   $\dot M=\dot M_{\rm out}/\dot M_{\rm Edd} = 0.5\pm 0.06$  (or 50\% of Eddington)   than during  OBS1 ($\dot M_1=0.40\errUD{0.05}{0.06}$).  Conversely, it appears that OBS4 caught a different phase or streamline of the wind, characterized by the lowest velocity and  correspondingly the  lowest mass outflow rate ($\dot M_4 =0.19\errUD{0.03}{0.03}$).  

 \begin{figure}[t] 
\rotatebox{-90}{
\includegraphics[angle=0,width=0.33\textwidth]{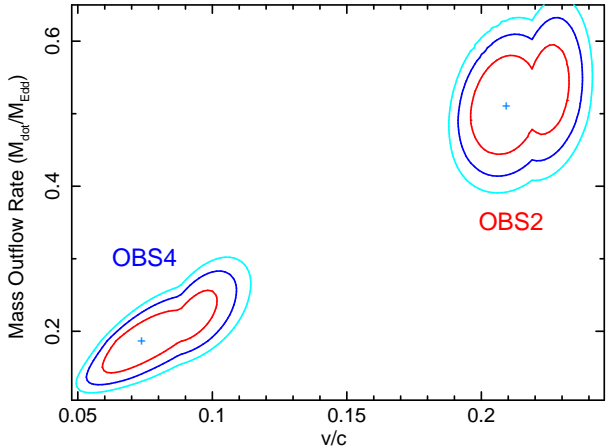} 
}
\caption{{Confidence contours for the  mass outflow rate versus the wind terminal velocity ($v/c$)  for OBS2  (upper right contours) and OBS4 (lower left contours).  The contours shown are at the   90\%,   99\% and 99.9\% significance levels for 2 interesting parameters. The best fit values are marked with   blue crosses. It is clear that the wind  parameters in  OBS2 and OBS4 occupy two distinct loci. 
}
 \label{fig:cont}
}
\end{figure}
To confirm that, we produced the contours of $\dot M$ versus the outflow velocity for OBS2 and OBS4, these are shown in Fig.~\ref{fig:cont}. It is noticeable that the wind caught in OBS4 is characterized by a much lower outflow rate and velocity than in OBS2, at a confidence greater than 99.9\%. 
We note that a lower mass outflow rate does not imply that the wind is characterized by a lower mass load;  indeed,  the low value measured during OBS4 can be  simply explained by the low  velocity  of the wind ($v_{\infty, 4}/c=-0.074 \errUD{0.014}{0.010}$), which is almost a factor 3 lower than the velocity measured in OBS2 ($v_{\infty, 2}/c=-0.210\errUD{0.010}{0.009}$). In support to this explanation, we can obtain a first order estimate of the column density of the flow along the line of sight by estimating  the factor by which the wind suppresses, by scattering, the 2-10 keV continuum. This in turn translates into the Compton depth of the wind  and therefore  into a first order  estimate of its column density. For OBS2, OBS3  OBS4, we found that the Compton depths    are similar  and   are of the order of $\tau \sim  0.9$  for a corresponding $\nhsym\sim 1.3 \times 10^{24}$\,\nh.    This confirms that the main driver  for the low $\dot M_4$, is the lower velocity of the wind. The Compton depth is only marginally lower in  OBS1, with  $\tau \sim  0.7$ and $\nhsym\sim   10^{24}$\,\nh,  and  may explain the slightly lower mass outflow rate. 

 It is worth noting that   the estimates of the  absorbing column densities are similar to the one derived with the photoionization modelling, where a $ \nhsym\sim 9.4 \times 10^{23}$\,\nh, was inferred with the variable ionization model. More importantly, the drop  in the outflow velocity in OBS4  does not depend on the assumed model, despite the  differences in the physical assumptions between the models. \\

    \begin{deluxetable*}{lcccc}[t]
 \tablecaption{Outflow Energetics Derived with the Disk Wind Model }
 \tablehead{& \colhead{OBS1}   & \colhead{OBS2}  & \colhead{OBS3} & \colhead{OBS4}} 
 \startdata
  \hline\\
$\dot{M}_{\rm out}$$^{a}$&$0.40\errUD{0.05}{0.06}$    & $0.51\errUD{0.05}{0.05}$  & $0.50\errUD{0.06}{0.06}$  & $0.19\errUD{0.03}{0.03}$         \\
 $v_{\infty}/c^{b} $ & 	      $-0.205\errUD{0.009}{0.009}$  & $-0.210\errUD{0.010}{0.009}$&$-0.194\errUD{0.009}{0.009}$  & $-0.074 \errUD{0.014}{0.010}$ \\
 $\% \dot{E}^c$&$9\pm 2$ & $11\pm 2$ & $9\pm 1$ & $0.5\pm 0.2$\\
$\dot{p}^{d}$ & $0.82\pm0.14$ & $1.07\pm 0.14$ &$0.97\pm0.15$ & $0.14\pm0.05$
 \enddata
 \tablenotetext{a}{Mass outflow rates in   Eddington units. }
\tablenotetext{b}{Mean wind velocity, see \S 5   and 6.1 for details. }
 \tablenotetext{c}{Percentage   outflow kinetic power in Eddington units. }
   \tablenotetext{d}{Outflow momentum rate in Eddington units.}
\label{diskwind_en}
\end{deluxetable*}

 \section{Discussion}\label{discussion}
  
 \subsection{The energetics of the disk wind}
 Here we compare the energetics  for  the disk wind  in \sorg\ that can be derived   with the \textsc{diskwind}   and with the \textsc{xstar} modelling.
 First of all,   from the \textsc{diskwind} model we derived that   the difference in the outflow rates  is driven by  changes of the outflow velocity rather than changes of  the mass load of the wind. Indeed,  the corresponding  column density is always of the order of $\nhsym \sim 10^{24}$\,\nh\ and  the variation in the wind opacity is explained by the changes of the ionizing X-ray luminosity ($\L_{\rm X}$), which sets the ionization of the wind.   
 Therefore, we  consider the energetics derived  with the \textsc{xstar}  model where we allowed the ionization rather than the $\nhsym$ to vary. 
 We also consider only the slow component of the wind, because  the \textsc{diskwind} model shows that the second zone could be an artifact of the modelling.
 We can follow the same argument described in  B21, where  the mass outflow rate normalized to the Eddington rate can be derived with the equation: 
 \begin{equation}   
\dot M=\frac{\dot M_{\rm {out}}}{ \dot M_{\rm {Edd}}}=2 \frac{\Omega}{4\pi} \mu_{\rm a} \,N_{\rm {H}}\,\sigma_{\rm {T}}\,\eta\,\left(\frac{v_{\rm{out}}}{c}\right)^{-1}
     \end{equation}  
   which   assumes a biconical geometry for the flow (\citealt{Krongold2007}), where  $\mu_{\rm a}=n_{\rm H}/n_{\rm e}=1.3$ for solar abundances, $\sigma_{\rm {T}}$ is the Thomson cross section and $\eta=0.1$ is the   accretion efficiency.  Here we also assume, following the same arguments presented in B21,  that  the wind subtends  a solid angle of $\Omega/4\pi = 0.5$  and  that the  launching  radius of the wind is the escape radius ($R_{\rm {min}}=2GM_{\rm {BH}}/v^2$), which gives the most conservative estimates of  the mass outflow rate and energetics  (see \citealt{Gofford2013,Tombesi2012}).

 In  OBS1, OBS2 and OBS3 the wind has $\nhsym= 9.4\errUD{1.4}{1.2}\times 10^{23}$\,\nh\ and  an averaged velocity of $v_{\rm {out}}/c = -0.148\pm 0.013$ (see Table~\ref{tab:xstar}); the derive mass outflow rate is  thus $\dot{M}_{\rm out\,1,\,2,\,3}/\dot{M}_{\rm Edd}=0.54\pm 0.09$. Conversely,  for OBS4 we derive  $\dot{M}_{\rm out, 4}/\dot{M}_{\rm Edd}=1.12\pm 0.23$, which is much higher than what is deduced from the \textsc{diskwind} fits, see below.
 
 The   kinetic power   of the wind ($ L_{\rm {KIN}}=1/2 \dot  M_{\rm {out}}\, v_{\rm{out}}^2$) normalized to the Eddington luminosity  ($L_{\rm {Edd}} =\eta\, \dot M_{\rm {Edd}} \,c^2$) is then:
\begin{equation}
\dot E= \frac{L_{\rm {KIN}}}{L_{\rm {Edd}}}=\frac{\Omega}{4\pi}\, \mu\, N_{\rm {H}}\,\sigma_{\rm {T}}\frac{v_{\rm{out}}}{c}
  \label{eq:E}  \end{equation} \\

 Using equation \ref{eq:E}, the kinetic power of the wind is then  $\dot E_{1,2,3}=6\pm 1$\% of Eddington in the first three observations and  $3\pm 1$\% in OBS4.  
  We note that the higher value of the mass outflow rate derived for OBS4 is driven by the  higher $R_{\rm {min}}$ ($\sim 5\times 10^{15}$\,cm), while the higher kinetic power derived for the first three observations is driven by the higher outflowing velocity. 
  
The kinetic power  and outflow rates are within the range of the measurements reported for the past observations (B18, M19, B21). In particular, the outflow rate and kinetic power measured in the first three observations are similar to the  energetics  measured  for the phase outflowing with $v\sim -0.2\,c$ that was detected in the past \swift\ and \suzaku\ observations (see B21). On the other hand the 
   mass outflow rate and kinetic power measured in OBS4 are close to the energetics measured for the slow component of the wind as seen  in the \suzaku\ observation, where we measured $ \dot{M}_{\rm out}/\dot{M}_{\rm Edd}\sim 0.84$  and $\dot E\sim 2.5$\%.  However, a strong assumption in these  and in the past estimates of the energetics of the wind is that the wind is launched at the escape radius, which may not be correct. Indeed, as we will discuss below,  it is implausible that the launching radius changes so dramatically, from $R_{\rm {min}}\sim 10^{15}$\,cm (or $\sim  50 R_{\rm {g}}$ as inferred for OBS1, OBS2 and OBS3) to $\sim 5\times 10^{15}$\,cm (or  $\sim 350 R_{\rm {g}}$  as measured in OBS4),  in $\sim 16$ days    and the most plausible scenario is that  the wind  always originates at   the same distance from the black hole. \\

    This limitation is not present in the diskwind model, where the wind streamline is launched from a fixed  innermost radius  $R_{\rm {min}}=64 R_{\rm g}$ and an outer radius of $1.5  R_{\rm {min}}$ and the terminal velocity of the wind does not have to be at the exact escape velocity at the innermost radius. In Table~\ref{diskwind_en}    we report the mass outflow rate,   terminal  velocity,     kinetic power and momentum rate of the wind for each of the observations.  We note that since the   wind  streamline has a physical extent,  determined by $R_{\rm {min}}$ and $R_{\rm {max}}$,  it also has    a range   of terminal velocities.  The terminal velocities and energetics reported in Table~\ref{diskwind_en} are derived for $R_{\rm {min}}$ ($v_\infty=f_{\rm v} \sqrt{2GM_{\rm BH}/R_{\rm {min}}}$). However, if we assume   the mean launch radius and its  mean velocity\footnote{This is the mean between the terminal velocity  at the inner launching radius  (i. e. $64 R_{\rm g}$) and the velocity at the outer radius of $96 R_{\rm g}$. },  the    kinetic power and momentum rate  would be only a factor of 20\% and 10\% lower, respectively.
We found that the mass outflow rate ranges between 40-50\% of Eddington for the first three observation, while it is lower in OBS4 and of the order of   $\dot M_4=19\pm3$\%. The lower value of $\dot M_4$ reflects the much lower velocity of the wind in OBS4. As we discussed in \S 5.1,   the difference in the derived mass outflow rates does not  imply  a different mass load of the wind  between the four observation, but it rather depends on the drop in velocity in OBS4 from $v_{\infty, 1,2,3}/c \sim  -0.2$  to $v_{\infty, 4}/c\sim -0.07$.  Indeed,  in \S 5.1 we showed that during the monitoring  the column density  of the flow along the line of sight is almost constant and of the order of  $ \sim 1.3 \times 10^{24}$\,\nh. The derived kinetic power  of the wind ranges between 0.5-10\% of Eddington, which for a black hole mass of $\sim 10^8 M_\odot$ (as estimated by B18), corresponds to $L_{\rm KIN}\sim 6\times 10^{43}-10^{45}$ erg s$^{-1}$.  Similarly to  the mass outflow rate results,   the lower value of $\dot{E_4}$ ($0.5\pm0.2$\%)  reflects the lower velocity of the wind in the last observation.\\

We can now compare these more robust estimates of the  disk wind kinematics  with the  energetics of  the possible kpc-scale molecular outflow  detected in the ALMA observation ($\dot E_{\rm CO}\sim 10^{42}$\, erg s$^{-1}$, $\dot p_{\rm CO}\sim 8\times 10^{34}$\, g cm s$^{-2}$; \citealt{Sirressi2019}). We confirm that we do not detect a boost of the momentum rate of the putative molecular outflow compared to the X-ray disk wind. Instead, the momentum rates of the two outflows are consistent. Furthermore, the kinetic power of the disk wind is still well above   the    kinetic  energy  carried  by  the   molecular gas phase. Precisely,  even if we   consider  the lowest   $\dot{E}$ of OBS4,  we still obtain that   $\dot E_{\rm CO}/\L_{\rm KIN}$ is     $\sim 0.02$.

  This result does not imply that the  disk wind in \sorg\ does not have an impact on the host galaxy, but it may simply reflect a lower efficiency  in coupling with the host galaxy gas (see a recent review by  \citealt{Veilleux2020}). Lower efficiencies in transferring the kinetic energy to the large scale molecular gas, like the one measured for \sorg, have  now been inferred for several powerful disk winds (\citealt{Mizumoto2019}), among them  some of the most   powerful disk winds, like  PDS\,456  and IZW1 (\citealt{Bischetti2019,ReevesBraito2019}). The lack of an energy conserving outflow can also be a sign that the central SMBH has not reached its critical M-$\sigma$ mass; in this case the momentum that can be imparted by the ultra fast disk wind is not enough to sweep the gas up to the cooling radius,  where the wind shock no longer cools. In this case the wind will not be able to expand adiabatically as in an energy  conserving outflow. The outflow will eventually fall back, because its momentum is not  enough to carry the weight of the gas that it has swept-up (see \citealt{King2003, King2005,King_Pounds2015}). 
  
  Critically, our analysis shows how challenging it  is to compare the energetics of the large scale outflows   and  the  ultra fast disk winds. 
Indeed, according to the   disk wind model,     the kinetic power of the wind in \sorg\ changes by more than an order of magnitude on a timescale as short as a month.  Therefore,  as already pointed out in other works (\citealt{Nardini2018,Zubovas2020}),  the  energetics derived with single epoch observations may be highly  inaccurate, because they may not reflect the typical  kinetic power of the wind. A single observation  determines the  disk wind properties at a specific time, while  disk winds  are generally extremely variable,  as demonstrated by  multi epoch observations of the best examples of winds  (e.g. PDS~456, \citealt{Reeves2018b,Matzeu2017}; IRASF\,11119+3257, \citealt{Tombesi2017}; PG\,1211+143, \citealt{pg1211} and  APM\,08279+5255, \citealt{Saez2011}).  A second, but no less important, aspect that could  have a profound impact when comparing these energetics is that, unlike with  \textsc{xstar},  the \textsc{diskwind} model provides a self consistent estimate of the energetics. For example,  the kinetic output  measured   for OBS4 with the \textsc{xstar} and  the \textsc{diskwind} model differs  by almost an order of magnitude.

  \subsection{The location  of the disk wind}
    We can now place some constraints on the location and sizescale of the streamline of the wind by means of its variability. The component of the disk wind that is  outflowing with $v_{\rm {out}}\sim  -0.2\,c$  lasts at least for   the  first three observations (see Table~\ref{tab:xstar} \&  \ref{tab:diskwind}), which   cover a timescale of 14 days (see Table~\ref{tab:obslog}). From this we can derive that the sizescale   of this absorber is $\Delta R=v\times \Delta T_1\sim 7\times 10^{15}$ cm (where  $\Delta T_1=1.2\times 10^6$\,sec).   From the column density of the wind derived  in \S 5.1 ($\nhsym \sim 10^{24}$\,\nh), we   can then estimate that its  density is of the order of  $n_{\rm e}= \nhsym/\Delta R \sim 10^8$ cm$^{-3}$.  
    
     From the photoionization modelling,  we derived that the ionization of the wind is of the order of log($\,\xi /{\rm erg\,cm \,s^{-1})}\sim 5$ (see Table~\ref{tab:xstar}); thus from the definition of the ionization parameter $\xi=L_{\rm {ion}}/n_{\rm e}R^2$ and assuming that the ionizing luminosity\footnote{ $L_{\rm {ion}}$ is the  ionizing luminosity over the 1-1000 Rydberg range.  Since \sorg\ is a highly obscured AGN, we cannot directly measure this  luminosity, we  therefore assumed as in B18 that it is of the order of 1/3 of the bolometric luminosity of $3 \times 10^{45}$ erg s$^{-1}$} is $10^{45}$ erg s$^{-1}$, we derive a typical distance from the central black hole of $R\sim 10^{16}$\,cm, or $\sim 700 R_{\rm g}$.    
     Thus, it appears  that the wind is located at a distance similar to its size (or $\Delta R/R\sim 1$), which supports a scenario where during this first three observations our line of sight intercepts a rather homogeneous streamline extending from the launch radius of $64 R_{\rm {g}} $ up to $\sim 10^3 R_{\rm {g}} $ and not a small clump or inhomogeneity of the wind.
           
 We now consider the slower phase of the wind detected in OBS4, which is outflowing at $v_{\rm {out}}\sim - 0.07\,c$. This slower zone or streamline of the wind emerges in OBS4, which occurred 16 days ($\Delta T_2=1.4\times 10^6$\,sec)  after OBS3.  Since we do not know  when this zone emerged,  following the same arguments outlined above,  we can derive an upper limit on its sizescale  of $\Delta R_{\rm {max}} < v_{\rm {out,4}} \times \Delta T_2 < 3 \times 10^{15}$ cm, which corresponds to $\sim 200 R_{\rm g}$. Likewise the density is $n_{\rm e}> \nhsym/\Delta R > 3\times  10^8$ cm$^{-3}$ and $R < 6\times 10^{15}$ cm, which is marginally lower than the location derived for the faster component of the wind. This in turn means that  in OBS4 our line of sight intercepts a slower   component of the wind  that could be actually located closer to the black hole than the faster one.     As we will discuss in \S 6.4, the possibility that  in OBS4 our line of sight intercepts a slower   component of the wind  that could be actually located closer to the black hole than the faster one  is in agreement with a  a recently proposed
scenario for the multiple velocities components observed in several winds.  According to this model   faster components are predicted to occur at larger radii after the inner and slower wind is accelerated through UV line driving (\citealt{Mizumoto2021}).\\

   \subsection{The extraordinary variability of the disk wind}
   Here we discuss what we can learn on the nature of the disk wind from the variability it  displayed in this monitoring. The key aspect  is that the  \sorg\ is an extremely variable source,  where both its intrinsic continuum and the wind  vary on  short timescales.  Our spectral analysis also demonstrates that the increase in the wind opacity and thus the depth of the absorbing structures is caused by a decrease in the ionization of the wind rather than an increase in the density of the wind.  Critically, this is the first case where we witnessed a dramatic variability of the wind outflowing velocity.  In particular, here the magnitude of the velocity changes, which is a factor of $\sim 3$, is much greater than those seen  to date in other variable disk winds (e. g. PDS\,456; \citealt{Matzeu2017},    IRAS  13224-3809; \citealt{Chartas2018,Parker2017,Pinto2018}  and APM\,08279+5255; \citealt{Saez2011}). We remark that, as shown in \S 5, this variation does not depend on the model adopted to fit the broad band X-ray data.    \\

      A    standard interpretation for the  variation of the wind velocity would be that during   OBS4 our line of sight intercepts a slower clump in the wind, which may be located further  out with respect to the faster streamline seen in the earlier observations.  However, as discussed above,  the rather small changes in the opacity/ionization of the wind   observed between the first three observations  are compatible with a relatively smooth  flow. Indeed, during this monitoring we did not observe any  rapid increase of the opacity of the wind, as expected from a clumpy wind. This is confirmed by  our estimates  of the size and location of the absorber whose ratio ($\Delta R/R$) is of order of unity.  It is thus unlikely that in OBS4 we are observing a  clump of the disk wind. 
   An alternative possibility is that during the last observation we observe a new slower streamline, which is launched further out, while the faster component outflowing at $\sim -0.2\,c $ gets disrupted.  Following the standard argument, where the launching radius corresponds to  the escape radius ($R_{\rm {min}}=2GM_{\rm {BH}}/v^2$), we would derive that the faster  and slower components are launched from $\sim 50\,R_{\rm g}$ and $\sim 350 \,R_{\rm g}$,  respectively. However, this is highly  implausible as this would require that in less than 16 days the launching  radius changes by an order of magnitude.  \\
 
 Instead the velocity changes may be due to a response of the wind  to the variability of the intrinsic emission of \sorg.  Recently,  a  correlation between the outflow velocity  of the disk wind  and the intrinsic X-ray luminosity has been reported for PDS456 (\citealt{Matzeu2017}), APM  08279+5255 (\citealt{Saez2011}) and   IRAS 13224-3809 (\citealt{Pinto2018,Chartas2018}), suggesting that these disk winds could be radiatively driven.  The correlations found for these winds can be expressed as $v_{\rm out} \propto L^ \alpha$, where the  $\alpha$ ranges from $\sim 0.25$, as measured for PDS\,456 to $\sim 0.4$ for IRAS 13224-3809.  We note that  for a radiatively driven wind the wind velocity is expected to be proportional to $L^{0.5}$. 
 
  In the case of \sorg, not only are the variations  more substantial than what seen in other winds, but the wind does not respond to the luminosity, whereby  the wind is slower in a bright  state (OBS4) and faster  in the faintest state (OBS2). One could argue that for \sorg\ we are witnessing a delay  in the response of the wind velocity to the luminosity, where  the low velocity measured in OBS4 is caused by the low intrinsic luminosity in OBS2. However,  from the above scaling relation,  in order to reproduce a factor $\sim 2.8$   velocity change, one would expect the luminosity to have changed by a factor of $\sim 8$, while the intrinsic X-ray luminosity in OBS2 is  only a factor of $\sim 2$ lower (see Table~\ref{tab:diskwind} and Fig. ~\ref{fig:nustar_lc}).  The only other possibility is that there was a more substantial decrease of the luminosity  in between OBS3 and OBS4;  however, this is highly unlikely as none of the previous observations caught \sorg\ in such a faint state (see Table 2 in B21).

\subsection{The Overall Scenario of the Wind in \sorg}
We now consider the  results   obtained by \citet{Mizumoto2021}    on the  X-ray   spectra   predicted from the most   recent hydrodynamics simulations of UV line driven disc winds, using a specific prescription for the X-ray opacity (\citealt{Nomura2020}). The above authors compute the evolution of the line driven wind properties (in terms of  ionization,  density and velocity)  as a function of the wind radius (e. g. see Fig. 5 of  \citealt{Mizumoto2021}). According to these simulations,  multiple velocities components are naturally produced in the flow, once acceleration due to  UV line driving is considered. In particular, they show that a highly ionized wind  that is launched at  a radius of about $\sim 100 R_{\rm g}$ with a velocity of $\sim 0.1\,c$, might be accompanied by a much faster component located at larger radii.  Here the inner highly ionized component  of the flow acts as the X-ray shielding gas, the so-called ``hitchhiking gas''  (\citealt{Murray1995}, see also \citealt{GiustiniProga2021}), which  prevents the outer wind from  becoming over-ionized. As the density of the wind increases with its radial distance, at about $\sim 500 R_{\rm g}$ its ionization state drops to  low enough (log$\xi <5$) to  produce the typical Fe-K absorption features.  Since this gas is  shielded, it can be accelerated  in situ by   line driving and reaches  a terminal velocity of the order of $0.25\,c$. The proposed scenario can thus naturally explain the different velocity components observed in several winds, which are not necessarily different streamlines, but the evolution of a single one.  
 Interestingly, the radii at which the acceleration occurs in the  \citet{Mizumoto2021} simulations  is the typical radius at  which we observe the absorbing gas in \sorg\ (i.e. of about $\sim 700 R_{\rm g} $ or $ \sim 10^{16}$\,cm).

\begin{figure}[t] 
\rotatebox{-90}{
  \includegraphics[angle=90,width=0.40\textwidth]{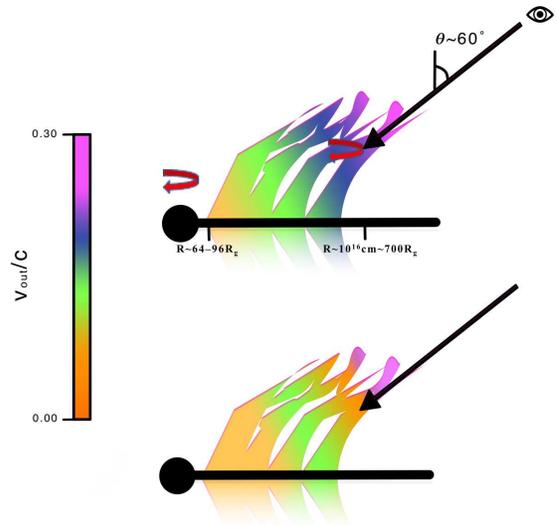} 
}
\caption{{A simple schematic of the wind streamlines, color-coded according to velocity -- from slow to fast -- orange, green, blue, violet (following the colors adopted by \citealt{Mizumoto2021}). The upper panel is representative of the first 3 observations, where our line of sight intercepts a fast wind component ($v_{\rm out} \sim -0.2\,c$), close to the likely terminal velocity. 
The red arrows represent the wind and black hole axis rotation. 
In the two weeks between observations 3 and 4, a new wind streamline emerges and the previous streamline has rotated out of the line of sight. A slower ($v_{\rm out} \sim -0.07\,c$) velocity component is intercepted (lower panel), which is yet to be accelerated to its terminal velocity. Approximate size scales for the launch radius and the wind streamlines are marked on the figure.  Note that the distances are not to scale. 
}
 \label{fig:scheme}
}
\end{figure}

According to  this scenario,    in the first three observations we may intercept  the flow after it is fully accelerated to $v_{\rm {out}}\sim -0.2\,c$. 
Conversely, during the last observation,  we  may be witnessing a new  streamline, which is yet to be fully accelerated. The lack of the faster phase of the wind in OBS4 can be explained  by simply considering the orbital timescale of the wind, which is much larger (i.e. greater than a year) than the dynamical timescale.
  Therefore, in the 16 days between OBS3 and OBS4 the accelerated streamline could have just simply rotated out of our  line of sight.  \\
A possible schematic for the wind in MCG-03-58-007 is shown in Fig.~\ref{fig:scheme}. 
This shows a simplified geometric representation of the possible wind streamlines seen across the observations, color coded by velocity. In the top panel, corresponding to OBS1-3, 
our line of sight intercepts the wind streamline,  which has been accelerated up to $v_{\rm out} \sim-0.2\,c$. 
In the lower panel (OBS4), a new streamline has emerged in the 16 days between observations. Here, the material 
has not yet been fully accelerated to its maximum terminal velocity and a lower velocity ($v_{\rm out}\sim-0.07\,c$) has been observed. Future, more intensive observations may be able to track further such velocity changes and how they occur, with a fine time sampling.  

 Looking back at the previous observations, these two phases were simultaneously in our line of sight in the \suzaku\ and possibly in the \swift\ observations (see B18 and B21). In particular, two deep absorption features were present in the \suzaku\ spectra at $\sim 7.4$\,keV and $\sim 8.5$\,keV (see B18). These structures   could be accounted for by  two highly ionised and high column density outflowing absorbers with  $v_{\rm{out1}}/c\sim -0.07$ and $v_{\rm{out2}}/c\sim -0.2$, matching the  two velocities measured  for the wind in this new monitoring campaign. 
Interestingly, in the past (during the long look with \xmm\ \& \nustar\ performed in 2015)  we also witnessed an occultation event,   where the opacity of the disk wind rapidly increased due to a possible  clump with a  higher than usual density (and consequently lower ionisation). From the duration of the occultation event, we derived a location for this higher  density and lower ionization clump of the wind of the order of $  R\sim 5\times 10^{15}$\, cm, which is consistent with the location of the wind derived  in this work.  An intriguing possibility is that these higher density clumps are indeed  the site of acceleration of the wind due to   line driving. We note that recent works  (\citealt{Dannen2020,Waters2021}) show  that AGN winds naturally become clumpy; although these works focus on thermally driven winds, a similar process may be at work in all winds.  As  winds evolve and propagate, the initial small inhomogeneities of the density or temperature of the gas will form  individual  denser/colder clumps or clouds, which can then  be the site of additional acceleration.

 \section{Summary and conclusion}
 
We presented the results of  a series of simultaneous \xmm\ and \nustar\ observations of \sorg\ performed in May-June 2019. During this campaign both the intrinsic emission and the  disk wind of  \sorg\  varied  on timescales as short as 4 days.        We confirm that the disk wind in \sorg\ is persistent, as it is detected in all the observations, but highly variable in both the opacity and velocity.  Indeed, both the \textsc{xstar} and the \textsc{diskwind} models  show that the variability of the disk wind opacity can be explained   by  changes in the ionization of the wind, which becomes less transparent during the second of the observations.   We also confirm that, regardless of the model adopted,  the   kinetic power  of    the disk windis  well above  the    kinetic  energy  carried  by  the   molecular gas phase. \\

   The novel and most striking result,  obtained with this new observational campaign, is that the outflowing velocity changed  by a factor of $\sim 3$ in just 16 days, dropping from $v/c\sim - 0.2$ to $v/c\sim -0.074$. To our knowledge this is the first case where such a dramatic variability of the wind velocity is observed. The  sizescale  and location of the absorber,   derived from  variability arguments,    suggest   that we are observing a rather homogeneous flow that is located at $\sim 700 R_{\rm g} $ (or $ \sim 10^{16}$\,cm)  and not  different small clumps of the wind. The  rapid variation is  the wind velocity is also incompatible with a scenario whereby we  intercept streamlines of the disk wind launched at different radii. Indeed, this  would imply that the launching radius changes from few tens  to few hundreds of $R_{\rm g}$ in just a couple of weeks, which is implausible.  A more likely scenario is that  we   observe the same streamline of the wind, but the  different velocity components   are  simply  due to the acceleration of the wind, as recently  proposed by \citet{Mizumoto2021}.  According to  this scenario,    in the first three observations we intercept  the flow after it is accelerated to $v_{\rm {out}}\sim - 0.2\,c$. Conversely,  during the last observation we   see only the   slower pre-accelerated streamline, while the accelerated phase  could have simply rotated out of our  line of sight.

Our results on the disk wind of \sorg\ emphasize the importance of monitoring programs of   disk winds.   A single snapshot can give an incomplete picture of them and   only observations at different timescales of the winds can    fully reveal  their complex structure.  Further monitoring programs of \sorg\ and similarly variable winds, are now clearly required to  better understand their  structure and  to  shed light on their driving mechanisms.   These,  combined with  ongoing
efforts  with the disk wind simulations,   will allow us to better understand  the evolution of the AGN winds and properly quantify the energy feedback exerted on the galactic environment.  

  \acknowledgements
 We thank the referee for his/her useful comments that improved the
paper.  We would like to thank Stuart Sim for the use of his disk wind radiative transfer code and for providing input into the   generation of  the diskwind  synthetic spectra used in this paper. This research has made use of data obtained from \xmm, an ESA science mission with instruments and contributions directly funded by ESA Member States and the USA (NASA), and from the \nustar\ mission, a project led by the California Institute of Technology, managed by the Jet Propulsion Laboratory, and funded by NASA. This research has made use of the NuSTAR Data Analysis Software (NuSTARDAS) jointly developed by the ASI Science Data Center and the California Institute of Technology.  VB acknowledges  financial  support  through   the NASA  grant   80NSSC20K0793.  VB, RDC, PS  acknowledge financial contribution from the agreements ASI-INAF n.2017-14-H.0. 
MG is supported by the ``Programa de Atracci\'on de Talento''  of the Comunidad de Madrid, grant number 2018-T1/TIC-11733.

 \software{xstar \citealt{xstar}), diskwind code (\citealt{Sim2008,Sim2010a,Sim2010b}), SAS (v16.0.0; \citealt{SAS}), FTOOLS (v6.27.2; \citealt{FTOOLS}), XSPEC (v12.11; \citealt{xspecref})}\\

\bibliography{biblio}
 \end{document}